\documentclass{aa} 
\usepackage{graphicx}
\usepackage{array}
\newcolumntype{L}{>{$}l<{$}}
\newcolumntype{R}{>{$}r<{$}}
\newcolumntype{C}{>{$}c<{$}}
\usepackage{physics}
\usepackage{hyperref}
\hypersetup{
    colorlinks=true,
    linkcolor=blue,
    citecolor=blue}
\usepackage{txfonts}
\usepackage{ulem}
\usepackage{xcolor}
\definecolor{blazeorange}{rgb}{1.0, 0.4, 0.0}
\definecolor{seagreen}{rgb}{0.18, 0.55, 0.34}
\definecolor{rufous}{rgb}{0.66, 0.11, 0.03}
\definecolor{royalfuchsia}{rgb}{0.79, 0.17, 0.57}
\definecolor{scarlet}{rgb}{1.0, 0.13, 0.0}
\definecolor{royalpurple}{rgb}{0.47, 0.32, 0.66}
\definecolor{darkblue}{rgb}{0, 0, 0.66}
\begin{document}

   \title{Numerical simulations of internal shocks in spherical geometry: hydrodynamics and prompt emission}


   \author{A. Charlet
          \inst{1,2}
          \and
          J. Granot
          \inst{1,3,4}
          \and
          P. Beniamini
          \inst{1,3,4}
          }

   \institute{Astrophysics Research Center of the Open University (ARCO), The Open University of Israel, 1 University Road, PO Box 808, Raanana 4353701, Israel
         \and
             Centre de Recherche Astrophysique de Lyon (CRAL), ENS de Lyon, UMR 5574, Université Claude Bernard Lyon 1, CNRS, Lyon 69007, France
        \and Department of Natural Sciences, The Open University of Israel, P.O Box 808, Ra’anana 4353701, Israel     
        \and Department of Physics, The George Washington University, 725 21st Street NW, Washington, DC 20052, USA}
        
   \date{Received December 09, 2024; accepted May 20, 2025}

 
  \abstract
   {Among the models used to explain the prompt emission of gamma-ray bursts (GRBs), internal shocks is a leading one. Its most basic ingredient is a collision between two cold shells of different Lorentz factors in an ultra-relativistic outflow, which forms a pair of shock fronts that accelerate electrons in their wake. In this model, key features of GRB prompt emission such as the doubly-broken power-law spectral shape arise naturally from the optically-thin synchrotron emission at both shock fronts.}
   {We investigate the internal shocks model as a mechanism for prompt emission based on a full hydrodynamical analytic derivation in planar geometry, extending this approach to spherical geometry using hydrodynamic simulations.}
   {We used the moving mesh relativistic hydrodynamics code \texttt{GAMMA} to study the collision of two ultra-relativistic cold shells of equal kinetic energy (and power). Using the built-in shock detection, we calculate the corresponding synchrotron emission by the relativistic electrons accelerated into a power-law energy distribution behind the shock, in the fast cooling regime.}
   {During the first dynamical time after the collision, the spherical effects cause the shock strength to decrease with radius. The observed peak frequency decreases faster than expected by other models in the rising part of the pulse, and the peak flux saturates even for moderately short pulses. This is likely caused by the very sharp edges of the shells in our model, while smoother edges will probably mitigate this effect. Our model traces the evolution of the peak frequency back to the source activity time scales.}
   {}


   \maketitle
%

\section{Introduction}
Relativistic outflows are common to astrophysical phenomena featuring accretion and/or explosions. Such outflows are observed as sources of bright non-thermal emission, indicating conversion of their kinetic energy into radiation. Internal shocks, collisions between parts of the outflow with different velocities, are one of the proposed dissipation mechanism in many astrophysical contexts. They were first introduced by \cite{rees1978m87} to explain the ``knots'' -- resolved inhomogeneities in the jet of galaxy M87, and were subsequently invoked to explain emission of radio-loud quasars \citep{spada2001internal}, microquasars \citep{kaiser2000internal, malzac2014spectral}, and GRBs \citep{rees1994unsteady, sari1997cosmological, kobayashi1997can, daigne1998gamma}. Internal shocks arise naturally when assuming inhomogeneities in the (proper) velocity of the outflow. In this scenario, a faster shell catches up with a slower one ejected at an earlier time by the central engine. The shells collide, and under the right conditions \citep[see e.g.][for a review]{pe2014energetic} two shock fronts will form: a forward shock (FS) propagating into the slow shell and a reverse shock (RS) propagating into the faster one. Particle acceleration is expected at these shock fronts, which consequently powers the emission from the outflow.

A common approach used to calculate the dissipation efficiency and calculate the emissions from internal shocks assumes plastic collisions: shells merge inelastically and continue propagating as a single shell \citep{daigne1998gamma, beloborodov2000efficiency, spada2000analysis, guetta2001efficiency, bovsnjak2009prompt, malzac2014spectral, bustamante2017multi, rudolph2022multiwavelength, rudolph2023multicollision}. The energy dissipation is associated to a FS if the Lorentz factor (LF) of the fused shell is closer to the fast shell, or to a RS otherwise. While providing a useful approximation that allowed us to reproduce a number of features associated with internal shocks, this approach crudely approximates the location of the shock front. Few studies focused on the shock physics \citep[e.g.][]{kino2004hydrodynamic, pe2017dynamical} but did not study the temporal evolution for a generic parameter space. Such a work was done in \cite{rahaman2024internal} (hereafter R24a), who provided an analytical framework for internal shocks between two cold, homogeneous, unmagnetized shells of arbitrary proper velocities in planar geometry.

This hydrodynamical framework is then completed with the parametrization from \cite{genet2009realistic} for the optically-thin synchrotron emission of a single shock front propagating over a range of radii. They found an analytical solution for the observed flux using integration over the equal arrival time surface (EATS) for a Band function spectral shape. \cite{rahaman2024prompt} (hereafter R24b) built their model on this basis, adding refinements such as the ratio between shock front LF and downstream fluid LF, and considering contribution from both shock fronts using the estimates obtained in R24a. This allowed them to self-consistently calculate the flux of a single collision at any observed frequency, which can then be used as a building block for full light curves and/or spectra of relativistic outflows. In this framework, the relative positions of the shock fronts and its dependency on the initial conditions are an important factor for the shape of the observed light curve, a feature washed away by the ballistic approach. While the choice of initial conditions in most papers exploring the internal shock model for the GRB prompt emission were leading to a very short-lived FS of negligible contribution to the total emission, the choice made in R24b led to the two fronts contributing over similar timescales. This allowed the FS to convert enough energy to have a sizable contribution to the emission, producing a two-component time-resolved spectrum. This physical scenario for prompt GRB emission may provide an explanation to the significant deviations from the pure Band function fit \citep{band1993batse} in the time-resolved analysis of the brightest bins of observed bursts \citep{preece2014first, vianello2018bright}.

The hydrodynamical solution is significantly simplified in the planar case, which is a decent approximation as long as the dissipation radius varies by a factor $\la 2$ during the shock propagation. In particular, the shock strengths remain fixed with time or radius in that case. Because of this R24a determined all quantities in planar geometry, and R24b introduced spherical effects on the peak frequencies and luminosities in an approximated manner by neglecting variation of the shock fronts LF and shock strength with time. Considering that such effects may modify the emission signature significantly, a fully consistent spherical approach is necessary to properly quantify the applicability of R24b results.

The present study focuses on the construction of the thin-shell synchrotron emission corresponding to the thin cooling layer behind the shock in the fast cooling regime of spherical colliding shells through numerical simulations. In Sect. \ref{sec1:main} we introduce the physical framework of this study, beginning with a summary of the main results from R24a (Sect. \ref{sec1:hydro}), following with a presentation of the numerical code (Sect. \ref{sec1:num}), the model used to derive observed flux from our simulation results (Sect. \ref{sec1:flux}), and finishing with the numerical setups for this study (Sect. \ref{sec1:setup}). We present the results in Sect. \ref{sec2:main}, starting with the comparison of our fiducial spherical run with the planar case (Sect. \ref{sec2:sph_fid}) before exploring the fully spherical regime over the doubling radius (Sect. \ref{sec2:sph_big}). Our calculated observed flux is presented in Sect. \ref{sec2:flux} before we study the behavior of its peaks (Sect. \ref{sec2:pks}), and then try to derive information on the source activity from a few selected GRBs using those results (Sect. \ref{sec2:nu2Rcr}). Finally we give first insights of cooling effects on the observed spectra through an approximated marginally fast cooling regime (Sect. \ref{sec2:mFC}). We conclude and discuss our results in Sect. \ref{sec:ccl}.


\section{Physical scenario and numerical methods}\label{sec1:main}
\subsection{Hydrodynamical framework}\label{sec1:hydro}
In R24a, the authors show that the collision of two homogeneous cold relativistic shells is determined by seven basic parameters presented Table \ref{tab:basicParams}: the time $t_\mathrm{off}$ between the ejection of the two shells, their proper speeds $(u_1, u_4)$, where $u\equiv\Gamma\beta$, initial radial widths $(\Delta_{0,1},\Delta_{0,4})$ and initial kinetic energies $(E_{\mathrm{k}0,1}, E_{\mathrm{k}0,4})$. Here and elsewhere, subscript 0 denotes the initial values of properties, i.e. at ejection or at the collision radius. Alternatively, the activity time $(t_{\mathrm{on},1}, t_{\mathrm{on},4})$ and the power $(L_1, L_4)$ of the source during the emission of the shells may be used in place of the shells width and kinetic energy for a set of parameters focused on the source activity. The two are easily related through $\Delta_{0,i}=\beta_\mathrm{i} ct_\mathrm{on,i}$ and $E_{\mathrm{k0,i}}=L_\mathrm{i}t_\mathrm{on,i}$, assuming constant jet power and outflow velocity within each ejection interval. In particular, the assumption of no velocity spread within a shell means that all fluid elements move at the same velocity and its width doesn't change during propagation. As shown in Table \ref{tab:derivParams}, those seven parameters can be combined into four derived parameters required to describe the post-collision shock hydrodynamics: the collision radius $R_0$, the shells' radial widths ratio $\chi$, the proper velocities ratio $a_\mathrm{u}$, and the proper density ratio $f$. We also introduce the collision time $t_0 = R_0/\beta_4c$ in the source frame, where we implicitly choose $t=0$ at the ejection of the front of shell 4. The shells are assumed to be cold, $p_1=p_4=0$, and the proper density of shell $i$ is obtained from
\begin{equation}
    \rho'_\mathrm{i} = \frac{1}{\Gamma_\mathrm{i}(\Gamma_\mathrm{i}-1)}\dfrac{E_{\mathrm{k0,i}}}{4\pi R_0^2\Delta_{0,i}c^2}\;,
\end{equation}
where primes denote quantities measured in the rest frames of the corresponding fluids. The collision of the two shells produces a pair of shocks: a reverse shock (RS) propagating into shell 4 and a forward shock (FS) propagating into shell 1. The two shocked regions, region 3 (shocked shell 4) and region 2 (shocked shell 1), are separated by a contact discontinuity (CD). At the collision radius, the use of the shock jump conditions together with pressure equality across the CD and the equation of state (EoS) in the shocked regions allow for an analytical derivation of all relevant hydrodynamical quantities. Values at the collision radius will be noted by the subscript 0.

In planar geometry, quantities are constant across a shocked shell and with propagation. From the constant shock front velocities one easily derives shock crossing times and radii, determining the emission timing properties. We give in Appendix \ref{app:planhydro} the main results from R24a. Those do not hold anymore in spherical geometry, prompting the present numerical study. 

\begin{table}
\centering
\caption{Physical parameters of the setup.}
\begin{tabular}{C l}
\hline\hline
\text{Symbol} & \multicolumn{1}{c}{Definition} \\\hline
E_{\mathrm{k0,i}} & Available kinetic energy in shell $i$ just before collision \\
\Delta_{0,\mathrm{i}} & Radial width of shell $i$ just before collision \\
u_\mathrm{i} & Proper speed of shell $i$ \\
t_\mathrm{off} & Time between ejection of the shells \\\hline
L_\mathrm{i} & Source power during the ejection of shell $i$\\
t_\mathrm{on,i} & Activity time of the source during ejection of shell $i$\\\hline
\end{tabular}
\label{tab:basicParams}
\end{table}

\begin{table}
\centering
\caption{List of derived parameters.}
\begin{tabular}{C l C}
\hline\hline
\text{Symbol} & \multicolumn{1}{c }{Definition} & \text{Expression} \\\hline
R_0 & Collision radius & \dfrac{\beta_1\beta_4 c t_\mathrm{off}}{\beta_4-\beta_1}\\
\chi & Radial width ratio & \dfrac{\Delta_{0,1}}{\Delta_{0,4}}\\
a_\mathrm{u} & Proper speed ratio & \dfrac{u_4}{u_1}>1\\
f & Proper density ratio & \dfrac{n'_4}{n'_1}=\chi\dfrac{E_{\mathrm{k}0,4}}{E_{\mathrm{k}0,1}}\dfrac{\Gamma_1(\Gamma_1-1)}{\Gamma_4(\Gamma_4-1)}\\\hline
\end{tabular}
\label{tab:derivParams}
\end{table}

\subsection{Numerical method}\label{sec1:num}
This study is performed with the code \texttt{GAMMA} \citep{ayache2022gamma} to solve the equations of special relativistic hydrodynamics (SRHD) in one dimension, using a finite-volume Godunov scheme, the HLLC \citep{mignone2005hllc} solver for relativistic hydrodynamics, piecewise linear spatial reconstruction, and following an arbitrary Lagrangian-Eulerian approach. This means \texttt{GAMMA} can compute fluxes for any interface velocity. The HLLC solver adds a calculation of the contact discontinuity (CD) wave speed to the two-wave HLL solver \citep{harten1983upstream}, the default behavior of \texttt{GAMMA} sets the interface velocity to that of the CD. We will use this default setting for all of this work, as a mesh moving with the flow's velocity evolves propagating shells over a wide range of scales using bigger time steps: in such a mesh the limiting velocity becomes the sound speed, while it is the flow speed in fixed-mesh approaches. Such a moving mesh also offers the added benefit of Lagrangian behavior of the cells like natural refinement of zones with high gradients, as the cell size will follow the compression of the fluid.

\texttt{GAMMA} offers the choice between several time integration methods, among which we chose the third-order Runge-Kunta. The time step is adaptive, based on a Courant-Friedrich-Lewy (CFL) condition \citep{courant1928partiellen}. To be consistent with the derivation of R24a, we implemented the Taub-Mathews equation of state \citep{mathews1971hydromagnetic} following \cite{mignone2007equation}. We used none of the adaptive mesh refinement methods present in \texttt{GAMMA} for this work to be able to properly identify cells from one timestep to the next and compare their properties. Finally, we use the shock detection algorithm introduced in \cite{zanotti2010electromagnetic} already implemented in \texttt{GAMMA}. We find setting the shock detecting threshold to 0.15 produces satisfactory detection of the two fronts across the simulation. \texttt{GAMMA} also contains a method to inject an electron distribution in shocked cells and let them evolve using a reformulation of the cooling equation into an advection equation, which we will not use in this work given our choice of of assumptions for the radiative efficiency \citep[see discussion in][and references therein]{rahaman2024prompt}. We leave the exploration of different cooling regimes to an upcoming work.

\subsection{Flux calculation from cells}\label{sec1:flux}
In this work we derive the observed flux from our simulations by applying the method for an infinitely thin shell as described in \cite{genet2009realistic} to cells right downstream of each shock front, the infinitely thin shell approximation is valid as $\Delta r/R\sim10^{-9}$ for shocked cells in the simulations run for this work. The contributing cells are chosen right downstream of the shocks, counting only one contribution of each cell crossed by the shock, waiting the few time steps necessary for the shock to cross to another cell to add a new contribution. We also assume the accelerated electrons follow a power-law energy distribution of index $p$ ($dN_e/d\gamma_e\propto\gamma_e^{-p}$ for $\gamma_e\geq\gamma_m$), and all the energy given to the electrons of a cell by the shock passage is radiated in less than a numerical time step. This requires the emission to be deep in the ``fast cooling'' regime. After identifying the contributing cell right downstream of the shock, we derive the minimum Lorentz factor of the post-shock electron distribution by normalizing the total available energy over the electron population:
\begin{equation}
    \gamma_\mathrm{m} = \frac{p-2}{p-1}\frac{m_\mathrm{p}}{m_\mathrm{e}}\frac{\epsilon_\mathrm{e}}{\xi_\mathrm{e}}\frac{e'_\mathrm{int}}{\rho' c^2}\;,
\end{equation}
for $p>2$, where $e'_\mathrm{int}$ is the comoving (or proper) internal energy density. $\epsilon_\mathrm{e}$ is the fraction of internal energy transferred to the fraction $\xi_\mathrm{e}$ of total electrons. For this study, we assume equipartition of energy $\epsilon_{\rm e}=\epsilon_{\rm B}=1/3$, and choose the values $p=2.5$, $\xi_{\rm e}=10^{-2}$ for the accelerated electron distribution power-law index and fraction of accelerated electrons respectively. To this Lorentz factor corresponds the comoving peak frequency
\begin{equation}
    \nu'_\mathrm{m}=\nu'_{\mathrm{syn}}(\gamma_\mathrm{m})=\frac{q_\mathrm{e} B'}{2\pi m_\mathrm{e} c}\gamma_\mathrm{m}^2\;.
\end{equation}
The contribution of a cell at radius $r$ and time $t$ traveling with dimensionless velocity $\beta$ (corresponding to the Lorentz factor $\Gamma$) to the flux at an observed frequency $\nu$ and observed time $T$ is:
\begin{equation}\label{eqn:Fnu_GG09}
    F_{\nu}(\tau\geq1)=\frac{1+z}{4\pi d_\mathrm{L}^2}\tilde{L}_{\nu_\mathrm{m}} \tau^{-2} S\left(\tau\frac{\nu}{\nu_\mathrm{m}}\right)\;,
\end{equation}
where $S$ is a normalized function verifying $S(x)=xS(x)=1$ for $x=1$. The normalized time $\tau$ is defined as
\begin{equation}
    \tau = \frac{T-T_\mathrm{ej}}{T_\theta},\quad\quad T_\mathrm{ej}=(1+z)\left(t-\frac{r}{\beta c}\right),
    \quad T_\theta = \frac{(1+z)r}{2\Gamma^2c},\label{eqn:obsT}
\end{equation}
and the peak luminosity:
\begin{equation}
    \tilde{L}_{\nu_\mathrm{m}} = \frac{\Gamma\Delta V^{(3)} }{T_\theta} \frac{\epsilon_\mathrm{e} e'_{\rm int}}{W(p)\nu'_\mathrm{m}},\label{eqn:localLum}
\end{equation}
with $\Delta V^{(3)}$ the (three-)volume of the cell in the rest frame of the central source, and $W(p)=2(p-1)/(p-2)$. Eq.~(\ref{eqn:localLum}) is obtained by comparing the isotropic energy from a single pulse in the formulation of GG09 and the formulation of \cite{de2012gamma}, the derivation is detailed in Appendix \ref{app:flux_norm}. The quantity $\tilde{L}_{\nu_\mathrm{m}}$ is a numerical equivalent luminosity derived from synchrotron power and is not to be confused with the usual luminosity that appears in similar equations for the flux in GG09 or R24b. From their Appendix D, we write the comoving luminosity behind a shock front\footnote{Since the electrons are assumed by construction to be deep in the fast cooling regime, it follows that the bolometric luminosity is directly related to $\nu_{\rm m}$ and the spectral luminosity at $\nu_{\rm m}$ and is independent of the cooling frequency.}:
\begin{equation}
    L'_{\nu'_{\rm m}} = \frac{L'_{\rm bol}}{W(p)\nu'_{\rm m}} =  \frac{4}{3}\pi R^2 \beta_{\rm ud}c \frac{\epsilon_{\rm e} e'_{\rm int}}{W(p)\nu'_{\rm m}}\;,
\end{equation}
using $e'_{\rm int} = (\Gamma_{\rm ud} -1)\rho_{\rm d} c^2$ to obtain $\beta_{\rm ud}$. Finally we identify :
\begin{equation}
    \tilde{L}_{\nu_\mathrm{m}} = \frac{3\Gamma \Delta r}{ \beta_{\rm ud}cT_\theta}L'_{\nu'_\mathrm{m}}.
\end{equation}
In this work, the spectral shape $S$ will either be the synchrotron broken power-law with low- and high-frequency spectral slopes $b_1$ and $b_2$ respectively (syn-BPL), or the Band function:
\begin{equation}
S(x) = e^{1+b1} \begin{cases}
    x^{b_1}e^{-(1+b_1)x}\quad&x\leq x_\mathrm{b}\\
    x^{b_2}x_\mathrm{b}^{b_1-b_2}e^{-(b1-b2)}\quad&x\geq x_\mathrm{b}
    \end{cases}\label{eqn:defBand}
\end{equation}
where $x_\mathrm{b}=(b_1-b_2)/(1+b_1)$. Following R24b, the choice of setting both fronts deep in the fast cooling regimes sets the values $b_1=-1/2$ and $b_2=-p/2=-1.25$ for spectral slopes. This choice will be discussed in the conclusion. The effect of setting the weaker FS in the marginally fast cooling regime will be discussed in Sect. \ref{sec2:mFC}. Summing all contributions from the run gives the total observed flux as a function of observed time and frequency.

In the following, quantities will be plotted in units of the normalization frequency $\nu_0$ and flux $F_0$, defined as the peak observed frequency and flux radiated at the RS at collision in R24b:
\begin{align}
    \nu_0 &= 2 \Gamma_0 \nu'_\mathrm{m,RS}(R_0),\\
    F_0 &= \frac{(1+z)}{12\pi d_\mathrm{L}^2}2\Gamma_0 L'_{\nu'_\mathrm{m},\mathrm{RS}}(R_0).
\end{align}
$\Gamma_0$ is the Lorentz factor of the shocked material at $R_0$, which is at this radius equal downstream of both shock fronts. See their Appendix G for the full derivation. We will write $\tilde{\nu}=\nu/\nu_0$ the normalized frequency, and $\tilde{T}=1+\bar{T}=(T-T_{\rm ej,0})/T_{\rm\theta,0}$ the normalized time as defined in Eq.~(\ref{eqn:obsT}) with the relevant values at $t_0$ and $R_0$.

\subsection{Numerical setup}\label{sec1:setup}
To expand on the theoretical framework of R24b, we explore the collision of two ultrarelativistic shells of equal energy and width with a moderate proper velocity ratio $a_\mathrm{u}=2$. The kinetic energy available in the shells is taken to be $10^{52}$ erg, a typical order of magnitude value for the isotropic equivalent value corresponding to a single GRB pulse. The choice of equal activity time / equal shell width is doubly motivated: by GRB spectrum observations and to ensure high radiative efficiency. From R24b, the peaks' relative prominence is tied to the shells' sizes: if $t_{\rm on,1}\gtrsim 2\,t_{\rm on 4}$ the FS peak is too prominent relative to the RS and vice versa (see e.g. their Fig. 4). They also showed that the rarefaction wave traveling back towards the center after shock crossing may catch up with and suppress the second shock before it finishes crossing its shell for values of $\chi$ not close to 1, limiting the radiative efficiency of the process in such a case. The off time between pulses $t_\mathrm{off}$ is set to the same value as the pulses themselves, as observations suggests a correlation between pulse width and interval between pulses, outside of quiescent periods \citep[see e.g.][]{nakar2002time}. We set both activity times $t_\mathrm{on}$ and the off time to a typical order of magnitude for the activity timescale of 0.1 s. The shells are ultra-relativistic with Lorentz factors of $u_1=100$ and $u_4=200$, a minimum value to ensure the assumption of optically thin emission.

To complete this setup we define the external medium density behind shell 4, $\rho'_\mathrm{ext,b}$ and in front of shell 1, $\rho'_\mathrm{ext,f}$  from a density contrast parameter $\tilde{\rho}'_\mathrm{ext}=\rho'_4/\rho'_\mathrm{ext,b}=\rho'_1/\rho'_\mathrm{ext,f}$. Both regions of external medium are set with the same velocity as the shell they are in contact with. Such an ``external'' medium is required for the rarefaction wave to set after shock crossing and hardly affects the results. The external pressure is defined by introducing the relativistic temperature $\Theta_0$ as a parameter, we set pressure to be constant everywhere to $p_0=\min(\rho'_1,\rho'_4)\Theta_0c^2$. Setting pressure equality over the whole simulation box avoids any unwanted effects from $pdV$ work between the external medium and the shells. All these parameters are summarized in Table \ref{tab:setup}. The simulation is run in 1D over 6100 cells of which $N_\mathrm{sh}=3000$ for each shell and 50 on each side for the external medium. This choice of resolution was made to ensure that the shock fronts have not propagated more than $2\times10^{-3}R_0$ before the downstream values are properly established. This can be estimated by applying the formula for the crossing time (Eq.~(\ref{eqn:t_sh})) to the width of 5 numerical cells, greater than the typical numerical size of the detected shock front with our choice of threshold parameter. Finally, we set an additional passive tracer with different values in each shell to help separate them easily during post-processing.

To fully explore the effects of spherical geometry on shock hydrodynamics by going well above the doubling radius for the crossing radius of each front, we ran another simulation in spherical geometry with larger activity times (and initial shells width) by a factor of 5 while keeping the off time constant. We increase the number of numerical cells per shell by the same factor. The values corresponding to this run are given in brackets in Table \ref{tab:setup}.

\begin{table}
\centering
\caption{Run parameters in CGS units.}
\begin{tabular}{C L}
\hline\hline
\text{Parameter} & \text{Value} \\\hline
E_{\mathrm{k},0} & 10^{52}\text{ erg}\\
u_1 & 100 \\
u_4 & 200\\
t_\mathrm{on,i} & 0.1\ (0.5)\text{ s} \\
t_\mathrm{off} & 0.1\text{ s}\\
N_\mathrm{sh} & 3000\ (15000)\\\hline
\rho'_1 & 4.7\times10^{-12}\text{ g.cm}^{-3}\\
\rho'_4 & 1.2\times10^{-12}\text{ g.cm}^{-3}\\
\Theta_0 & 5\times10^{-5}\\
\tilde{\rho}'_\mathrm{ext} & 5\times10^{-2}\\
\Delta_{0,i} & 3\times10^9 (1.5\times10^{10})\text{ cm}\\
t_0 & 2.7\times10^3\text{ s}\\
R_0 & 8\times10^{13}\text{ cm}\\\hline
a_\mathrm{u} & 2\\
\chi & 1\\
f & 0.25\\\hline
\end{tabular}
\label{tab:setup}
\end{table}

\section{Effects of spherical geometry}\label{sec2:main}
Having calibrated our simulation against the analytical results in planar geometry in Appendix \ref{app:planhydro}, we explore the spherical effects through two simulations: one where all parameters are equal for the sake of comparison, and a second one where activity times $t_\mathrm{on,i}$ and thus shell width $\Delta_{0,i}$ have been multiplied by 5 to explore hydrodynamics over a larger range of radii.

\subsection{Spherical effects on shells structure and shocks dynamics}\label{sec2:sph_fid}
\begin{figure*}
\centering
\includegraphics[width=.32\textwidth]{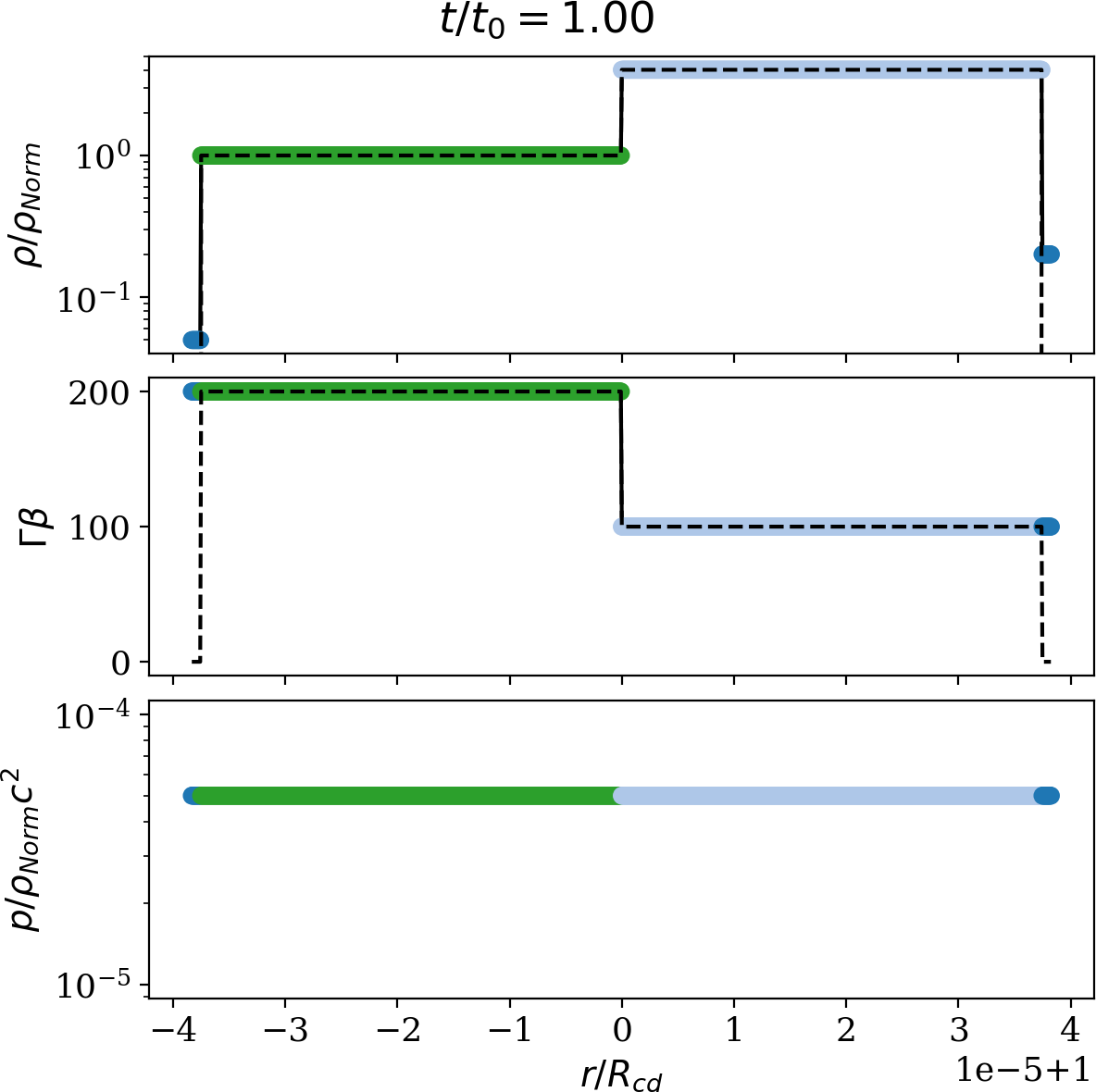} \includegraphics[width=.32\textwidth]{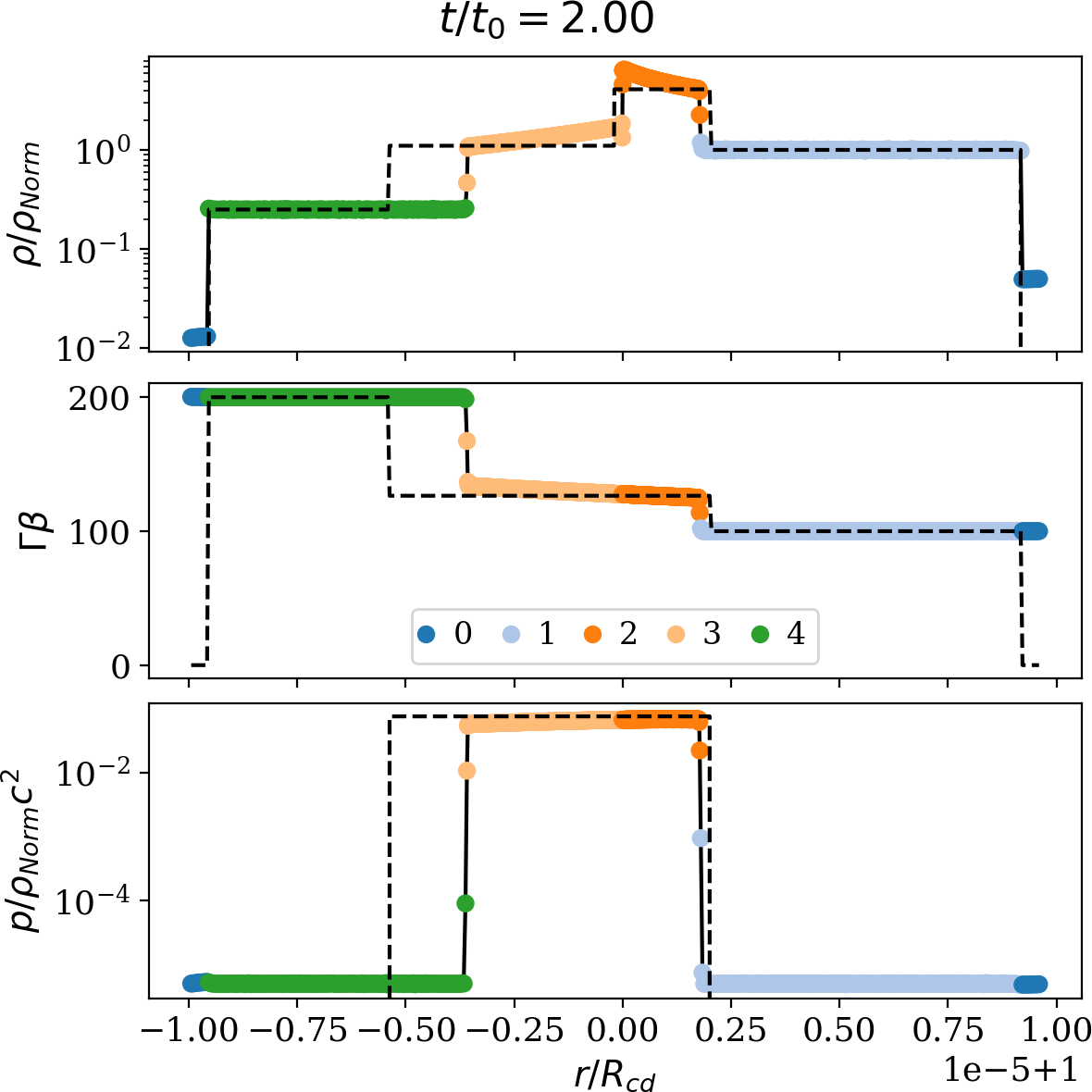} \includegraphics[width=.32\textwidth]{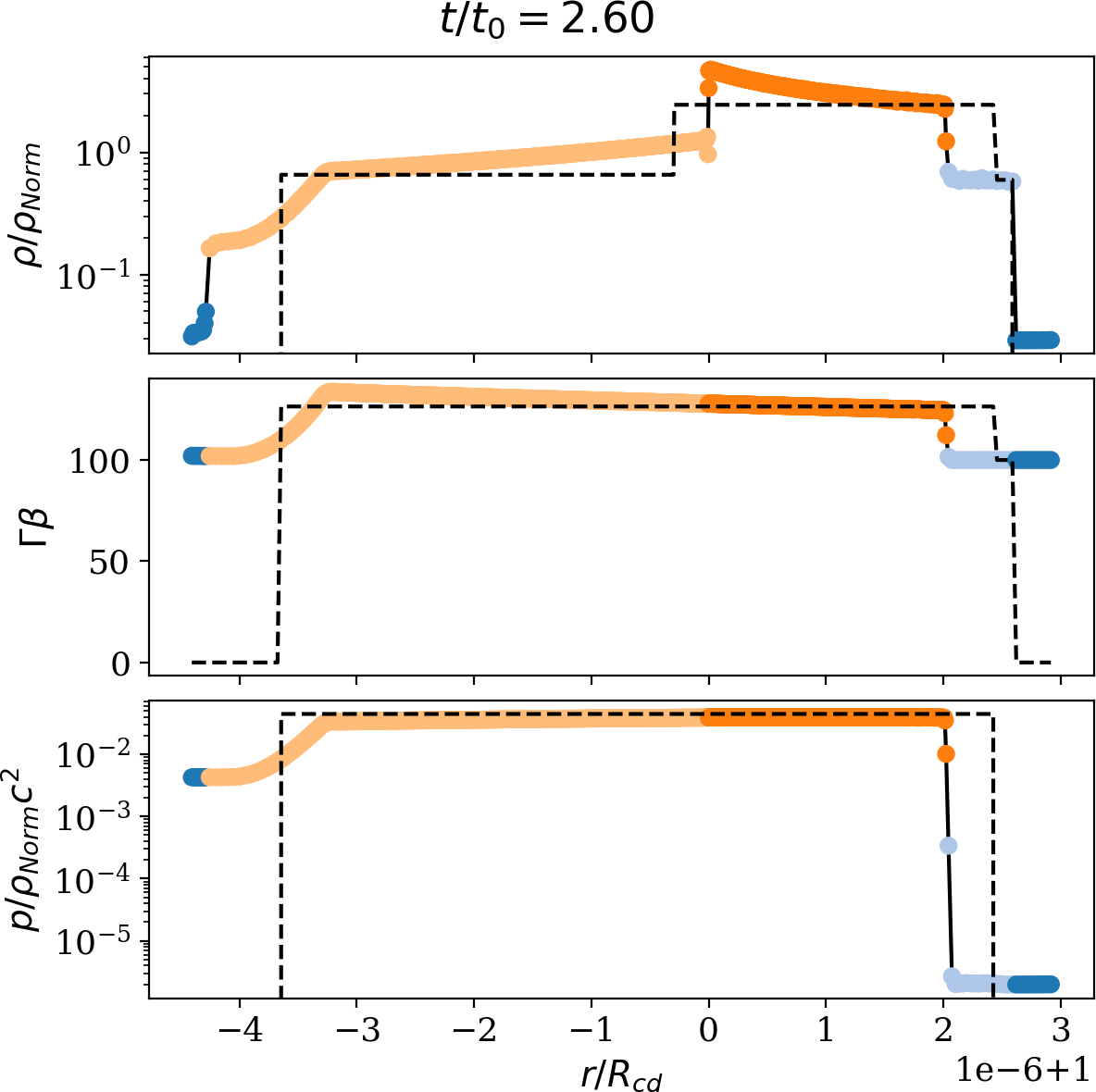}
\caption{Snapshots of the run in spherical geometry, with all parameters similar to the fiducial run, at $t=t_0$ (left), before first (reverse) shock crossing (center), and after shock crossing (right). Dashed black lines show the analytical expectations from the results in planar geometry. Legend is common to all panels.}
\label{fig:sph_snaps}
\end{figure*}

A first result of performing the same run in spherical geometry is the variation in crossing times (radii): for the same values of parameters, $\Delta R_\mathrm{RS}/R_0$ increases from 1.29 to 1.52 when switching from planar to spherical geometry, while $\Delta R_\mathrm{FS}/R_0$ only grows from 1.62 to 1.67. Fig. \ref{fig:sph_snaps} shows snapshots of the comoving density, proper velocity and pressure, similar to Fig. \ref{fig:cart_snaps}, at times $t_0$, 2 $t_0$, and 2.6 $t_0$. The analytical expectations from planar geometry, with density and pressure rescaled by $(R/R_0)^{-2}$ to account for propagation, are plotted to better highlight the differences with R24a. Spherical effects modify the structure of the shocked regions: the density profile shows compression from the shocks to the contact discontinuity, a proper velocity decrease with radius, and a slightly increasing pressure profile from the RS to the FS. As the shells and shocks propagate, the density in regions 1 and 4 (i.e. the unshocked portions of both shells) will decrease with $r^{-2}$. This is due to mass conservation of a fluid element in spherical geometry, with our assumption of little to no velocity spread between the two interfaces. At the 0th order where we consider constant shock strength, the downstream density $\rho_\mathrm{d}$ and pressure will also evolve following a $r^{-2}$ law. From pressure continuity (i.e. the pressure gradient is rather small in the shocked regions as they are in causal contact), the pressure at the CD will follow the same law instead of the $r^{-2{\hat{\gamma}}}$ from adiabatic expansion, with $\hat{\gamma}$ being the adiabatic index. That is, $p_d\approx p_0(R/R_0)^{-2}\approx p_{\rm CD}\approx(\rho_{\rm CD}/\rho_0)^{\hat{\gamma}}$ and $\rho_d=\rho_0(R/R_0)^{-2}$, implying $\rho_{\rm CD}/\rho_d=(R/R_0)^{2(1-\hat{\gamma}^{-1})}$.  This results in a compression of the fluid towards the center with the ratio between densities behind the shock and at the CD following a $-2(1-\hat{\gamma}^{-1})$ law. This implies a density scaling $\rho_{\rm CD}\propto r^{-2/\hat{\gamma}}$ between $r^{-1.5}$ and $r^{-1.2}$ ($\rho_\mathrm{CD}/\rho_\mathrm{d}\propto r^{2(1-\hat{\gamma}^{-1})}$ between $r^{0.5}$ and $r^{0.8}$) for an adiabatic index varying between $4/3$ and $5/3$.

\begin{figure}
    \centering
    \resizebox{\hsize}{!}{\includegraphics[width=\linewidth]{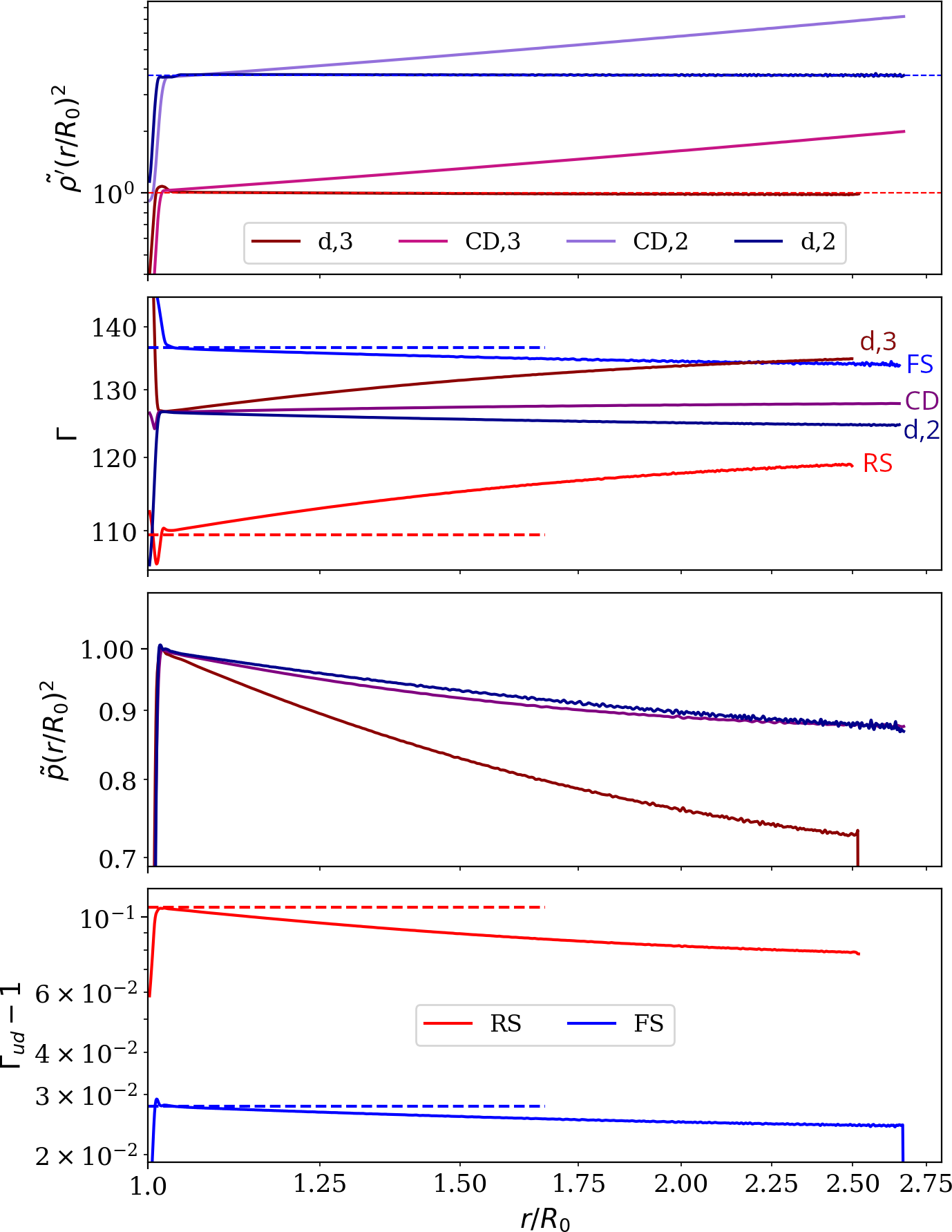}}
    \caption{From top to bottom: comoving densities, Lorentz factors of the fluid and shock fronts, pressure, shock strengths. The hydrodynamical quantities are measured downstream of shocks ('d,3' downstream of the RS, 'd,2' for the FS), at the CD by averaging over a few cells crossing the interface, or on each side of it (CD$_3$, CD$_2$) for the density. Legend is common to all panels. The derived Lorentz factor of the shock fronts is displayed with the measured ones. Densities and pressures are in units of the values behind the RS in the planar case and rescaled to $(r/R_0)^2$. We smoothed numerical oscillations in data by a rolling average. Horizontal dashed lines show the corresponding analytical values from R24a.}
    \label{fig:sph_hydro}
\end{figure}

Both of these effects are seen Fig. \ref{fig:sph_hydro}, top panel, which displays densities rescaled with $(r/R_0)^2$ to take out spherical effects: the values taken right downstream of the shocks show very little to no variation in radius, while those on both sides of the CD increase with the same $r^{0.7}$ (or  $\rho'\propto r^{-1.3}$ without rescaling) fitting with our observed regime of intermediate shock strengths. This results in intermediate adiabatic index values according to the \cite{kumar2003evolution} formulation used in R24a:
\begin{equation}
    \hat{\gamma} = \frac{4\Gamma_\mathrm{ud}+1}{3\Gamma_\mathrm{ud}}\;,
\end{equation}
which they proved to be equivalent to the Taub-Mathews EoS for a cold upstream medium. To a highest order of precision, the variation of the shock strength $\Gamma_\mathrm{ud}-1$ with propagation changes this behavior: as seen in the bottommost panel of Fig. \ref{fig:sph_hydro} the shock strengths decreases with propagation, at a stronger rate for the RS than the FS. This translates to a slight decrease of the post-RS rescaled density, and stronger decrease in post-RS pressure than downstream of the FS as shown in the 3rd panel of Fig. \ref{fig:sph_hydro}, explaining the increasing pressure profile from RS to FS seen in the snapshots. Similarly, we observe larger post-RS flow velocity than downstream of the FS, the first increases with propagation while the second decreases with it, creating the profile observed Fig. \ref{fig:sph_snaps}. While this approach in terms of power-law scalings proves itself useful to explain behavior below the doubling radius (i.e. $R\la 2R_0$), exploring a greater range of radii show different behavior above it.

\subsection{Asymptotical hydrodynamics in the spherical regime}\label{sec2:sph_big}
\begin{figure}
    \centering
    \resizebox{\hsize}{!}{\includegraphics[width=\linewidth]{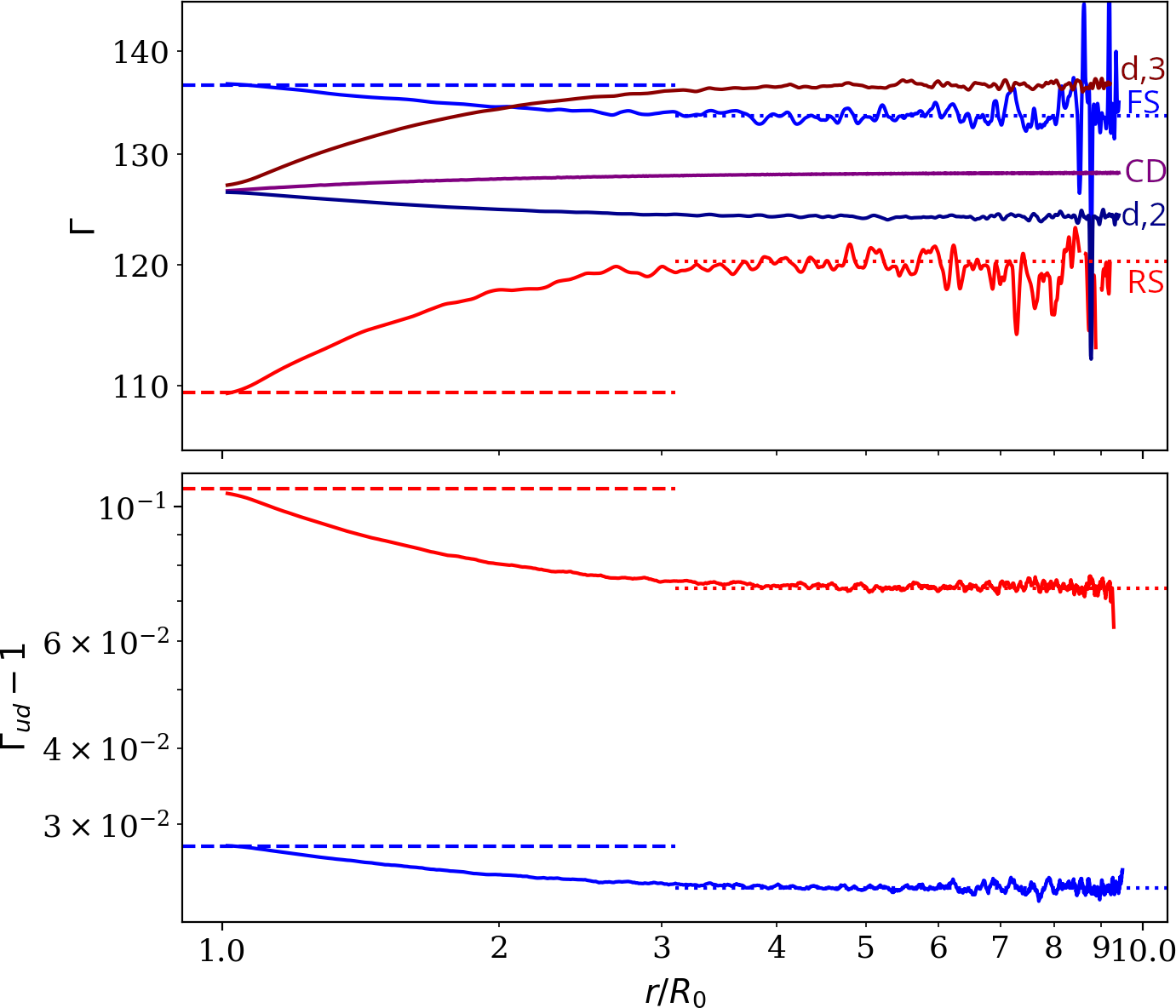}}
    \caption{Top: Lorentz factors of the shock fronts, the fluid downstream of the shocks and at the CD with radius. Bottom: Evolution of shock strengths with radius. Data was smoothed by rolling average. The quantities converge smoothly to a constant value after $r\approx4R_0$, shown by the dotted horizontal line for the quantities relative to the RS and the FS. Legend is common to Fig. \ref{fig:sph_hydro}.}
    \label{fig:sph_bigLF}
\end{figure}

We present here the results of the simulation ran with both activity times multiplied by a factor of 5, while keeping $t_\mathrm{off}$ and all other quantities constant. This means the shells are now wider than their separation by this same factor, allowing us to explore the evolution of the various relevant quantities for the flux calculation over a range of radii $\sim\!10\,R_0$, almost an order of magnitude greater than the situation presented in Sect. \ref{sec2:sph_fid}. In Fig.~\ref{fig:sph_bigLF} we show the fluid and shock fronts Lorentz factors in the top panel, and the shock strengths in the bottom panel. Data has been smoothed by a rolling average window to eliminate the strong noise caused by the propagation of numerical errors in the interface velocity over the large number of code iterations for this simulation. Such oscillations could be reduced by the use of an higher spatial order reconstruction algorithm instead of the 1st order piecewise linear algorithm currently present in \texttt{GAMMA}. The downstream and shock Lorentz factors evolve at the same rate, keeping the ratio $g=\Gamma/\Gamma_{\rm sh}$ introduced in R24a close to constant. At the collision radius the values are in accordance to the analytical expectations in planar geometry derived in R24a, before decreasing in a power-law up to the doubling radius and smoothly connecting to a constant value after a few $R_0$. Values relative to the RS change by a greater amount in this asymptotical regime compared to the planar case than values relating to the FS: shock and downstream Lorentz factor grow by $\sim10$ \% at the RS and go down by $\sim 3$ \% at the FS, while shock strength diminishes by $\sim3$ \% and $\sim0.7$ \% respectively. We thus model any hydrodynamical or derived quantity $X$ with the law:
\begin{equation}\label{eqn:smth_sph}
    X(R) = \left[\left(X_{0*}\tilde{R}^n\right)^s + \left(X_\mathrm{sph}\tilde{R}^h\right)^s \right]^{1/s} = X_0 f_X(R),
\end{equation}
with $\tilde{R}=R/R_0$ and $f_X$ a function verifying $f_X(R_0)=1$. The index $h$ corresponds to the expected scaling from propagation effects at constant shock strength, which can be 0; $X_\mathrm{sph}\tilde{R}^h$ is the asymptotical value at large radii, and $X_{0*}$ is defined by the value at collision radius $X_0 = X(R_0)=[X_{0*}^s+X_{\rm sph}^s]^{1/s}$. Note that $s$ need to be of the opposite sign of $n$ (and of $h$ when nonzero). The values obtained for the shock strength and downstream Lorentz factors by the fitting procedure are given Table \ref{tab:fit_hydro}. While the change in those quantities and thus the related quantities such as luminosity and peak frequency are not very significant in absolute values, it is their rate of change that has the most impact on observable data at very early times. 

\subsection{Emission from two spherical colliding shells}\label{sec2:flux}
\begin{figure}
    \centering
    \resizebox{\hsize}{!}{\includegraphics[width=\linewidth]{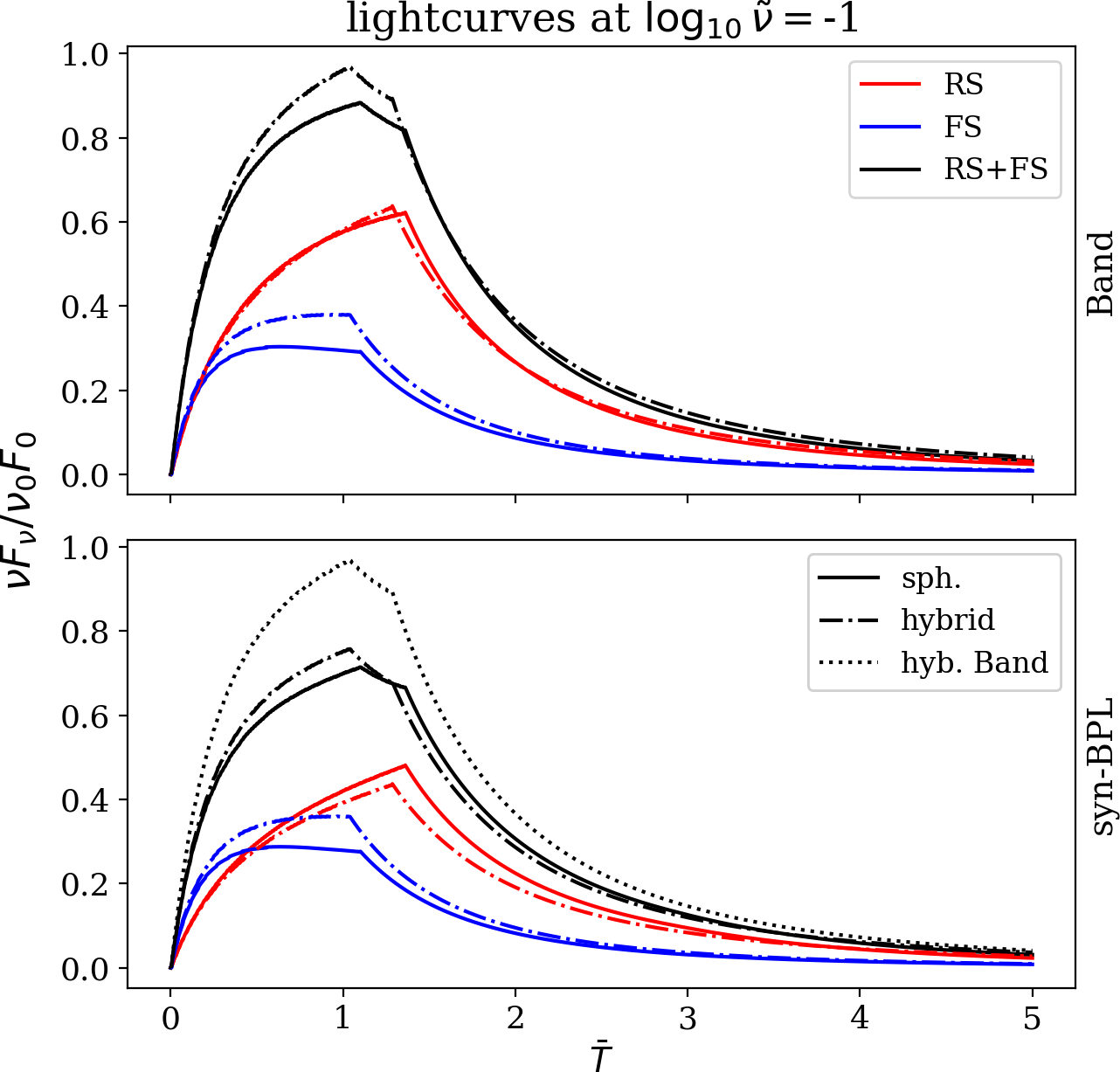}}
    \caption{light curves in the spherical (full lines) and hybrid (planar geometry and spherical assumptions for peak frequency and luminosity, dash-dotted lines) cases at fixed frequency $\tilde{\nu}=0.1$. Top panel shows the light curves with Band spectral shape and bottom panel the broken power-law spectral shape. The total light curve in the hybrid + Band case is redrawn in the bottom panel (dotted line) for the sake of comparison. The individual contributions from each shock change but the resulting light curve do not, apart from the slightly increased breaks from the increase in crossing times. The choice of spectral shape decreases total emission at this frequency while keeping the same time evolution.}
    \label{fig:comp_lc}
\end{figure}

\begin{figure}
    \centering
    \resizebox{\hsize}{!}{\includegraphics[width=\linewidth]{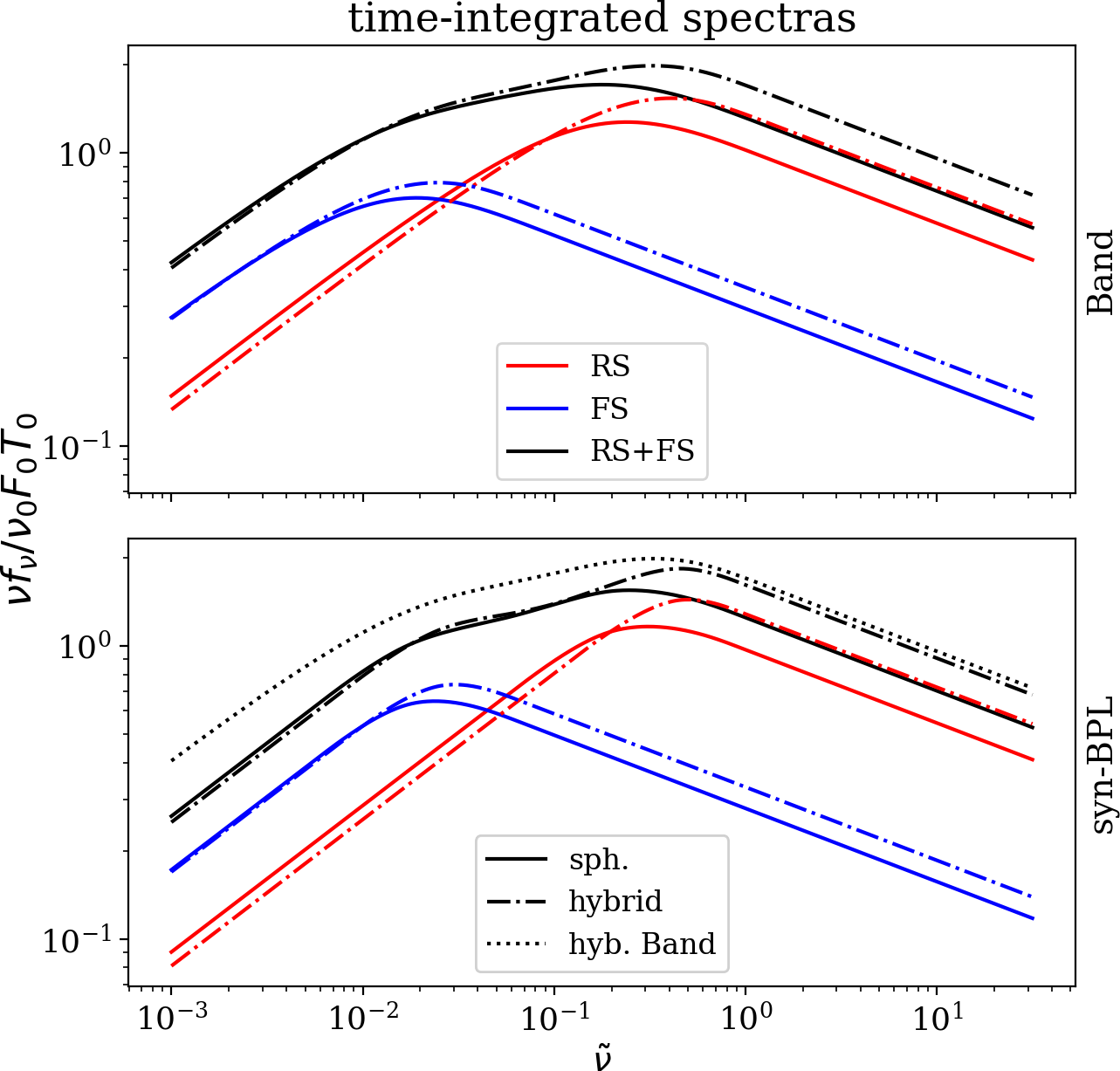}}
    \caption{Time-integrated spectra following the same structure as in Fig. \ref{fig:comp_lc}. The spherical effects decrease the peak frequencies for both shocks, with a stronger effect on the RS. The peaks from both contributions are closer than in the planar case, resulting in a three-part spectrum. This effect is more visible with the broken power-law spectral shape where the two peaks are distinct from each other in the hybrid approach. The choice in spectral shape also diminishes the flux at low frequency.}
    \label{fig:comp_intsp}
\end{figure}

We present in Figs. \ref{fig:comp_lc} and \ref{fig:comp_intsp} the resulting light curves and time-integrated spectra from our simulations, comparing the two geometries in the same panel for both the Band and the broken power-law spectral shape. To highlight the difference between the assumptions for propagation chosen in R24b and the fully spherical results, we compare our results to the hybrid approach instead of the fully planar case detailed in Appendix \ref{app:flux_calib}. The total light curve is weakly affected by the geometry, with slightly increased peak times corresponding to our estimation in Sect. \ref{sec2:sph_fid} and a smoothed shape compared to the planar case. In comparison, the individual contributions of each shock front vary significantly, with the RS peak luminosity increasing by a factor $5/4$ and the FS's decreasing by a factor $0.7$.

The time-integrated spectra in Fig. \ref{fig:comp_intsp} show the peak frequencies of both shock fronts decrease by different amounts: the quantities relative to the RS decrease faster with radius than those relative to the FS. Thus the observed range of the frequencies ratio between the two spectral components can be attained with a wider range of $a_\mathrm{u}$ in this fully spherical approach. We retain from the hybrid approach the doubly broken power-law spectrum with indices $1+b_1=0.5$ and $1+b_2=-0.25$ at low and high energies, directly related to our choice of synchrotron spectral slopes, and an intermediate part $1+b_{\rm mid}\approx0.1$. The effective middle slope $b_{\rm mid}$ depends on the flux normalization for each front and is thus very variable with the choice of hydrodynamical conditions: an analytical estimate using R24b show $1+b_{\rm mid}\in[-0.05,0.45]$ when exploring the parameter space $(\chi,a_{\rm u})\in[-0.5,1.5]\times[1.01,5]$, under the assumptions of ultra-relativistic shells and constant central engine power. These slopes are the same for both the Band and syn-BPL spectral shapes. The differences between the results of this present study and those from R24b are highlighted by redrawing in dotted line the Band + hybrid light curve (respectively time-integrated spectrum) in the syn-BPL panel. The main change lies in the total received flux (respectively fluence), especially at lower frequencies, but the essential characteristics of the light curve (respectively time-integrated spectrum) are similar. 

\subsection{Peak frequency and flux}\label{sec2:pks}
\begin{figure}
    \centering
    \resizebox{\hsize}{!}{\includegraphics[width=\linewidth]{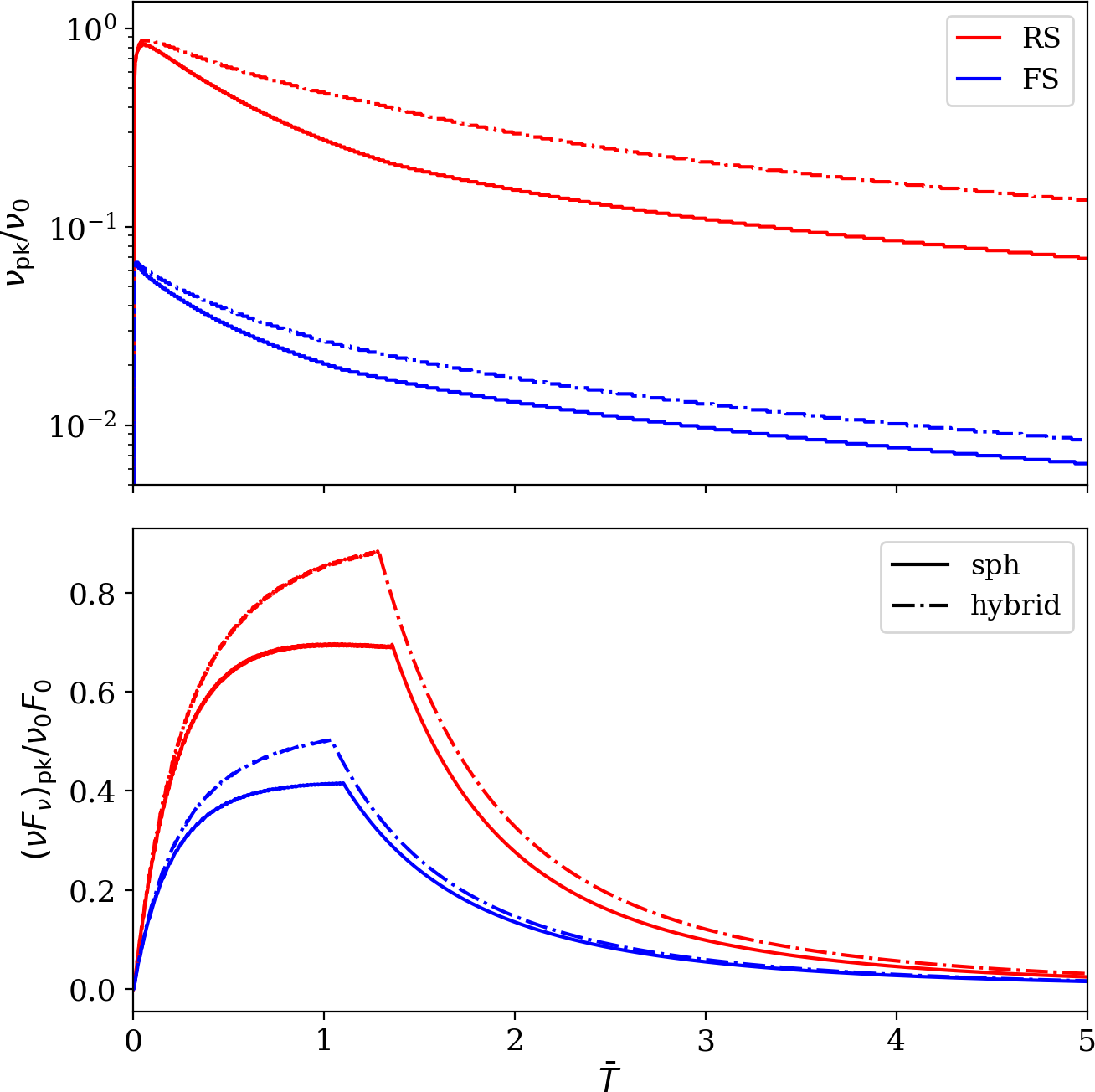}}
    \caption{The normalized time evolution of peak frequency and flux for both shock fronts in the spherical (full line) and hybrid (dash-dotted) cases, calculated from the flux obtained with the Band spectral shape. In spherical geometry, the peak frequency decrease at a faster rate during the rising phase and the flux reaches saturation before the break to the high-latitude emission.}
    \label{fig:pks_withT}
\end{figure}

The individual variations of peak frequency and flux associated to each shock front are displayed in Fig. \ref{fig:pks_withT} against normalized time, where the spherical effects systematically lowers both peak frequency and flux for both shocks. The spherical geometry also introduces a difference in the time evolution of the peak frequency between the rising and the high latitude emission (HLE) part. While the flux rises, the peak frequency decreases faster than $\nu_{\rm pk}\propto \tilde{T}^{-1}$ expected from purely geometric considerations. It is especially the case for the RS, with $\nu_\mathrm{pk,RS}\propto\tilde{T}^{-1.9}$. In the following we present a way to estimate those quantities from the hydrodynamics with less calculations than the full flux calculation detailed in Sect. \ref{sec1:flux}. We build this approach on the asymptotical regimes identified for the hydrodynamical quantities in Sect. \ref{sec2:sph_big}.

\begin{figure}
    \centering
    \resizebox{\hsize}{!}{\includegraphics[width=\linewidth]{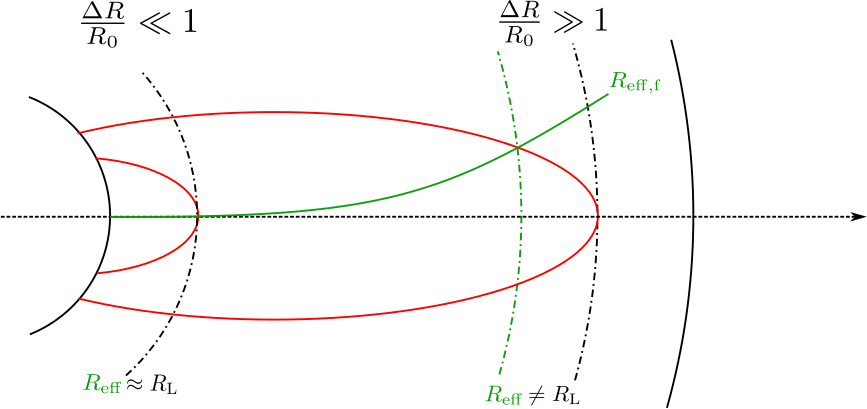}}
    \caption{Schematic representation of the effective radius approximation used to derive peak frequency and flux from the hydrodynamical data depending on the asymptotical regime. The EATS is shown in red.}
    \label{fig:xieff}
\end{figure}

Assuming the hydrodynamical variables can be modeled according to the law given by Eq.~(\ref{eqn:smth_sph}), after extracting the relevant variables for flux calculation from the simulation results we fit them with the same type of law :
\begin{align}
    \Gamma_\mathrm{d}(R) &= \left[\left(\Gamma_{0*}\tilde{R}^{-m/2}\right)^s+(\Gamma_\mathrm{sph})^s\right]^{1/s}\equiv\Gamma_{\rm f,0}f_\Gamma(R)\;,\\
    \nu'_\mathrm{m}(R) &= \left[\left(\nu'_{0*}\tilde{R}^d\right)^s+(\nu'_\mathrm{sph}\tilde{R}^{-1})^s\right]^{1/s}\equiv\nu'_{\rm f,0}f_{\nu'}(R)\;,\\
    L'_{\nu'_\mathrm{m}}(R) &= \left[\left(L'_{0*}\tilde{R}^a\right)^s+(L'_\mathrm{sph}\tilde{R})^s\right]^{1/s}\equiv L'_{\rm f,0}f_{L'}(R)\;.
\end{align}
We perform the fit with the \texttt{curve\_fit} function from \texttt{SciPy}, which is based on a non-linear least square method. We give in Table \ref{tab:fit_hydro} the obtained values for the fits. The numerical values at $R_0$ differ slightly ($\la2\%$) from the analytical expectations, we note them with the subscript "${\rm f,0}$" to avoid confusion.
\begin{table}[]
\caption{Fitting parameters for emission-related quantities.}
\centering
\begin{tabular}{C C C C C}
\hline\hline
& X_{0*} & X_\mathrm{sph} & \text{index} & |s| \\\hline
(\Gamma_\mathrm{ud}-1)_\mathrm{RS} & 0.103 & 0.0735 & n = -0.64 & 5.0 \\
(\Gamma_\mathrm{d}/\Gamma_0)_\mathrm{RS} & 1.01 & 1.08 & m = -0.29 & 24.0 \\
(\nu'_\mathrm{m}/\nu'_0)_\mathrm{RS} & 0.72 & 0.40 & d = -3.46 & 1.0 \\
(L'_{\nu'_\mathrm{m}}/L'_0)_\mathrm{RS} & 1.03 & 1.43 & a = 1.84 & 3.1 \\\hline
(\Gamma_\mathrm{ud}-1)_\mathrm{FS} & 0.0278 & 0.0236 & n = -0.21 & 18.5 \\
(\Gamma_\mathrm{d})_\mathrm{FS} & 0.999 & 0.982 & m = 0.06 & 102.2 \\
(\nu'_\mathrm{m}/\nu'_0)_\mathrm{FS} & 0.40 & 0.65 & d = -2.67 & 1.0 \\
(L'_{\nu'_\mathrm{m}}/L'_0)_\mathrm{FS} & 1.05 & 1.16 & a = 1.39 & 5.7 \\\hline
\end{tabular}
\label{tab:fit_hydro}
\end{table}
Then to obtain an estimate for the peak frequency and flux in the rising part of the prompt emission, we assume we can define an effective radius on the EATS at which the peak contribution can be calculated $R_\mathrm{eff}(T)=y_\mathrm{eff}R_\mathrm{L}(T)$, with $R_\mathrm{L}(T)$ the maximal radius of the EATS (found along the line of sight). At a given observer time $T$, this radius can also be identified by its angle with the line of sight, or more conveniently by the quantity $\xi_\mathrm{eff}=(\Gamma\theta)_\mathrm{eff}^2$. A schematic of this method is given Fig. \ref{fig:xieff}.

Both quantities are joined by $y=\left[1+g^{-2}\xi\right]^{-1}$, and we take $R_\mathrm{L}=R_0\tilde{T}$ where we implicitly choose a constant $\Gamma_\mathrm{sh}$ to approximate the EATS at the 0th order. Then the peak frequency and flux are given by:
\begin{align}
    (1+z)\nu_\mathrm{pk}(T) &= k_\nu \nu_{\rm f,0}\frac{f_\Gamma(R_\mathrm{eff}) f_{\nu'}(R_\mathrm{eff})}{1+\xi_\mathrm{eff}},\\
    F_{\nu_\mathrm{pk}}(T) &= k_F F_{\rm f,0} 3g^2 \xi_\mathrm{max}\left(\dfrac{f_\Gamma(R_\mathrm{eff})}{1+\xi_\mathrm{eff}}\right)^3\dfrac{f_{L'}(R_\mathrm{eff})}{f_\Gamma(R_\mathrm{L})^2}.
\end{align}
Where $\nu_{\rm f,0} = 2\Gamma_{\rm f,0}\nu'_{\rm f,0}$ and $F_{\rm f,0}= \left((1+z)/(4\pi d_{\rm L}^2)\right)2\Gamma_{\rm f,0}L'_{\rm f,0}$. $k_\nu$ and $k_\mathrm{F}$ are two free parameters of order unity introduced to compensate for the difference between analytical and numerical values at $R_0$ such that $k_\nu \nu_{\rm f,0}=\nu_0$ and $k_F F_{\rm f,0} = F_0$. The observed peak frequency is obtained by calculating the Doppler factor and comoving peak frequency at $R_\mathrm{eff}$, and the flux is obtained by assuming the flux integral (Eq.~(\ref{eqn:Fnu_th_basic})) can be approximated by the value of the integrand at $\xi_\mathrm{eff}$ times the extension of the EATS in this variable, $\xi_\mathrm{max}=g^2(\tilde{T}-1)$. We then look for the function $\xi_\mathrm{eff}(T)$ fitting the data. Noting that contributions from regions of the EATS beyond the relativistic beaming angle $1/\Gamma$ become increasingly suppressed, we define a saturation angle $\xi_\mathrm{sat}$ of order unity and we chose the fitting function for $\xi_\mathrm{eff}$:
\begin{equation}
    \xi_\mathrm{eff}(T) = k_\xi\left(\xi_\mathrm{max}^{\,-s}+\xi_\mathrm{sat}^{-s}\right)^{\,-1/s} = k_\xi\left(\left[g^2(\tilde{T}-1)\right]^{-s} + \xi_\mathrm{sat}^{\,-s}\right)^{-1/s},
\end{equation}
where $k_\xi$, $\xi_\mathrm{sat}$ and $s$ are the free parameters. The high-latitude emission part is modeled as the emission from the final effective radius $R_\mathrm{eff,f}=y_\mathrm{eff}(T_\mathrm{f})R_\mathrm{f}$, with $T_\mathrm{f}$ the observed time corresponding to the shock crossing:
\begin{align}
    \nu_\mathrm{pk,HLE}(T) &= \nu_\mathrm{pk}(T_\mathrm{f})\tau_{\mathrm{f}}^{d_\nu}\;,\\
    F_{\nu_\mathrm{pk,HLE}}(T) &= F_{\nu_\mathrm{pk}}(T_\mathrm{f})\tau_{\mathrm{f}}^{d_\mathrm{F}}\;,
\end{align}
and $\tau_\mathrm{f}$ is the normalized time defined at $R_\mathrm{eff,f}$ following Eq.~(\ref{eqn:obsT}). We give Table \ref{tab:fit_pks} the values found to fit the peak frequency and flux of Fig. \ref{fig:pks_withT} with an accuracy within a few percent.

\begin{table}[]
\centering
\caption{Fitting parameters for the peak frequency and flux.}
\begin{tabular}{c C C C C C C C}
\hline\hline
front & k_\xi& \xi_\mathrm{sat} & s & k_\nu & k_\mathrm{F} & d_\nu & d_\mathrm{F}\\\hline
RS & 0.29 & 6.98 & 1.0 & 0.98 & 0.54 & -1.31 & -2.76 \\
FS & 0.51 & 2.89 & 1.0 & 0.97 & 1.28 & -1.02 & -2.47 \\\hline
\end{tabular}
\label{tab:fit_pks}
\end{table}

\subsection{Maximal frequency and crossing radius}\label{sec2:nu2Rcr}
\begin{figure}
    \centering
    \resizebox{\hsize}{!}{\includegraphics[width=\linewidth]{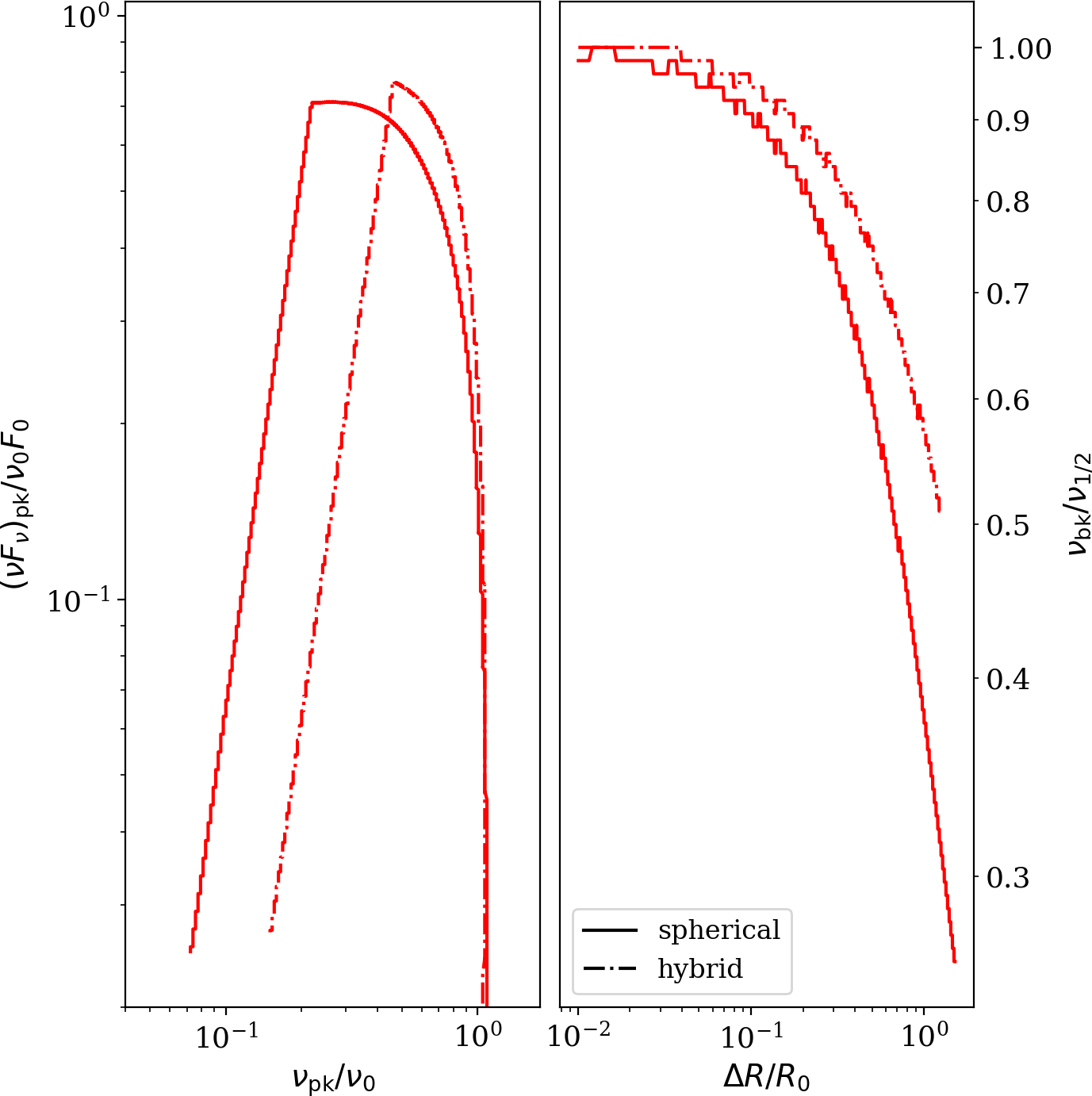}}
    \caption{Left: $\nu F_\nu$ at peak frequency vs peak frequency for the reverse shock in the spherical and hybrid case. The fully spherical case show peak flux over a larger range of frequencies. Right: Ratio between break frequency $\nu_\mathrm{bk}$ and frequency at half flux $\nu_{1/2}$. Legend is common to both panels.}
    \label{fig:comp_bhvNnubk}
\end{figure}

Among the spectral signatures, the behavior of the peak flux versus peak frequency shown left panel of Fig. \ref{fig:comp_bhvNnubk} is the one presenting the most striking change between planar and spherical dynamics. In this figure, we compare the purely spherical case to the hybrid approach of R24b, which features planar dynamics along with spherical geometry for calculating the radiation. The spherical case transitions to the anticipated correlation of high-latitude emission $(\nu F_\nu)_\mathrm{pk} \propto \nu_\mathrm{pk}^3$ at lower frequency but presents a high flux over a wider range. We will define $\nu_\mathrm{bk}$ as the peak frequency at this break. Comparing to Fig. I1 in R24a, the combined effects from spherical geometry on peak flux and frequency is greater than a pure increase of $\Delta R/R_0$ in the hybrid approach: simply increasing the crossing radius in the R24 formalism is not enough to obtain the curves in the spherical case.

In the right panel of Fig. \ref{fig:comp_bhvNnubk}, we display the evolution with normalized crossed radius $\Delta R/R_0$ of the ratio between $\nu_\mathrm{bk}$ and the frequency $\nu_{1/2}$, defined as where the peak flux in the rising part is half of that at $\nu_\mathrm{bk}$. This allows us to add our modifications to the results presented in Table I1 from R24b: from the ratio in frequencies for peak flux obtained from the data presented in \cite{yan2024one} and assuming the RS is the main contribution to the observed flux, they infer the radius crossed by the RS during the burst $\Delta R/R_0$ and thus the ratio of activity time to off time $t_{\mathrm{on},4}/t_\mathrm{off}$. We present in Table \ref{tab:nutodR} our updated values for the crossed radius in light of spherical effects and compare them to the values inferred in R24b. On average, the obtained crossed radius for our fiducial spherical case is half of the value obtained by R24b and about a third for the obtained activity over off time ratio. 

\begin{table}
\centering
\caption{Estimation of $\Delta R/R_0$ and corresponding $t_\mathrm{on}/t_\mathrm{off}$ for a choice of GRBs.}
\begin{tabular}{c C C C C C}
\hline\hline
GRB & \nu_\mathrm{bk}/\nu_{1/2} & (\Delta R/R_0)_{R24} & (\Delta R/R_0)_\mathrm{sph} & (t_\mathrm{on}/t_\mathrm{off})_\mathrm{sph}\\\hline
140606B & 0.45 & 1.66 & 0.81 & 0.57\\
131011A & 0.59 & 1.04 & 0.51 & 0.37\\
170607A & 0.36 & 2.36 & 1.09 & 0.74\\
151027A & 0.65 & 0.84 & 0.43 & 0.31\\
150514A & 0.60 & 0.99 & 0.50 & 0.36\\
120326A & 0.70 & 0.71 & 0.35 & 0.26\\
190829A & 0.50 & 1.40 & 0.69 & 0.49\\\hline
\end{tabular}
\label{tab:nutodR}
\end{table}

\subsection{Dissipation efficiency and marginally fast cooling regime}\label{sec2:mFC}
The total efficiency $\epsilon_{\rm tot}=\epsilon_{\rm int}\epsilon_{\rm e}\epsilon_{\rm rad}$ of the internal shock process is easily derived in this framework: being deep in the fast cooling regime implies a radiative efficiency $\epsilon_{\rm rad}=1$ behind both shock fronts. We note $\epsilon_{\rm int}$ the conversion rate of kinetic energy into internal energy, commonly called the dissipation efficiency (or ``thermal efficiency'', noted $\epsilon_{\rm th}$, in R24a,b). We calculate it by summing the internal energy from each contributing cell and comparing this quantity to the initial available kinetic energy. We obtain a dissipation efficiency of 8.5\% in the planar case, equal to the analytical value using Eq.~(B5) in R24b, and of 7.3\% in the spherical case. The total efficiency is thus $\epsilon_{\rm tot}=2.4$\% in the spherical case. In R24a,b the dissipation efficiency is obtained by comparing the total final internal energy present in the shells after shock crossing to the initial available kinetic energy. Both our ``local'' and their ``global'' approaches give the same result in the planar case, but quantities are different in the spherical case due to the shock strengths evolving as they cross their respective shells. Using their definition would give a dissipation efficiency of 3.5\% in the spherical case.

\begin{figure}
    \centering
    \includegraphics[width=\linewidth]{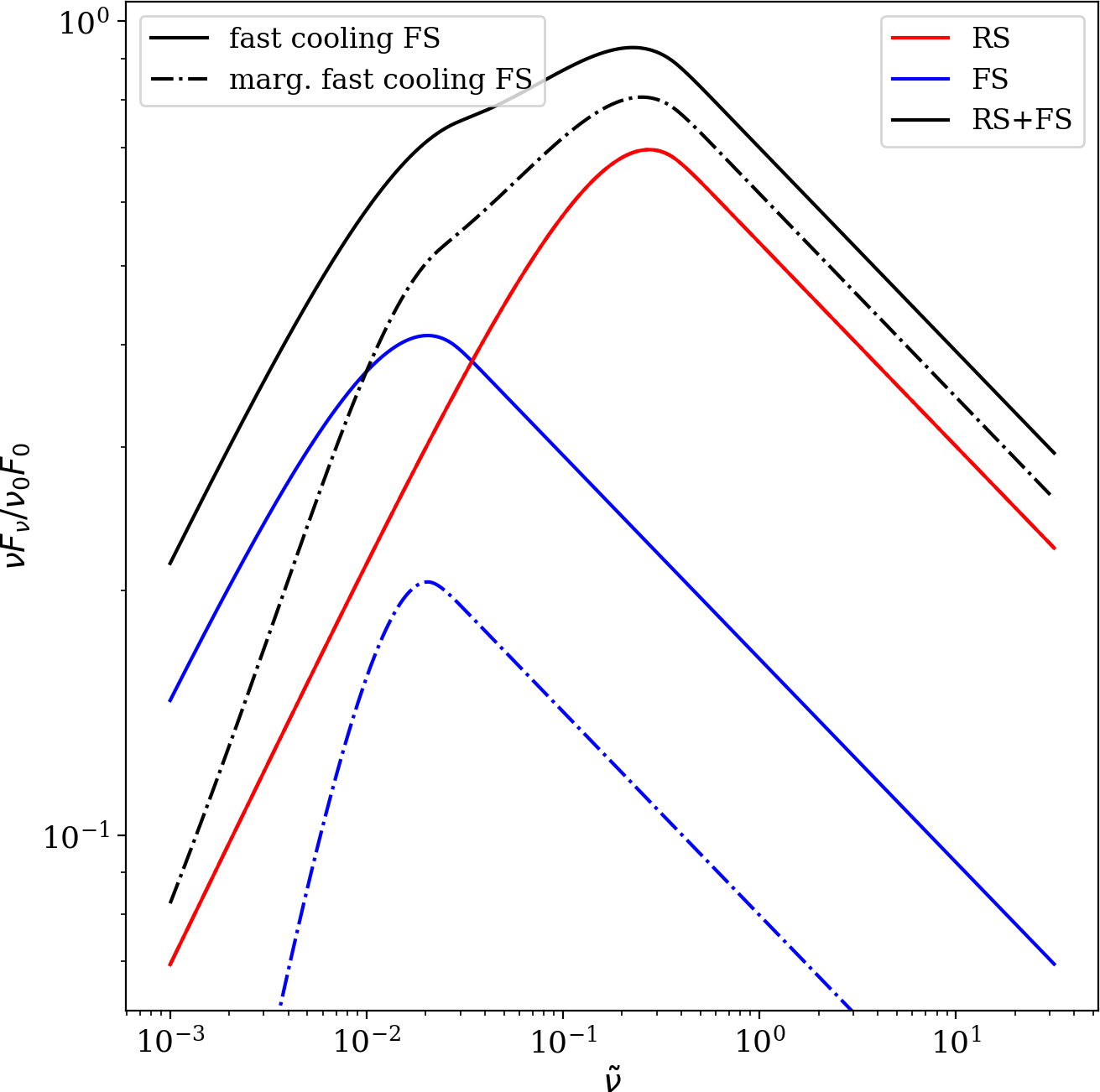}
    \caption{Comparison of instantaneous spectra between two fast-cooling shock fronts (full lines) versus a fast-cooling RS and a marginally fast cooling FS (dash-dotted lines) at $\bar{T}=1$. The low energy slopes are steeper and the overall flux is lower with a marginally fast cooling FS.}
    \label{fig:FCvmFC}
\end{figure}

The spectrum we obtain using this thin-shell assumption shows a low-energy $\nu F_\nu$ slope $1+b_1=0.5$, indicative from fast cooling, which does not seem to fit observed bursts. In e.g. \cite{ravasio2019evidence}, their fits using a doubly broken power-law model give values of the low-energy slopes from several bursts that are more consistent with a slow cooling regime. While a model where cooling is derived consistently is out of the scope of the current paper, we begin to look into this issue by recalculating the contributions from the FS in an approximated marginally fast cooling regime. For this we take $b_{\rm1,FS}=1/3$ instead of $-1/2$ and apply an efficiency factor $\epsilon_{\rm rad,FS}=1/2$ to the computed FS flux as in R24b. This value of 1/2 is an arbitrary middle ground between $\epsilon_{\rm rad, FC}\sim1$ in the fast cooling case and $\epsilon_{\rm rad, SC}\ll1$ in the slow cooling case. With this choice, we obtain an overall efficiency of the prompt emission $\epsilon_{\rm tot}=2$\% for the spherical simulation. We compare in Fig.~\ref{fig:FCvmFC} the instantaneous spectra at $T=T_0$ ($\bar{T}=1$) obtained in this marginally fast cooling FS case with the fast cooling regime. Both are calculated from the simulation in spherical geometry. For this marginally fast cooling FS we do not obtain the slow cooling low-energy slope of $1+b_{\rm 1,FS}=4/3$ but rather $\sim0.8$ from the cumulative contribution of the two shock fronts.

Setting the FS in the marginally fast cooling regime is not enough to obtain low energy slopes more in line with observations from our set of simulations. But one needs to mention that marginally fast cooling regime has three issues here. 1. It requires a very specific set of physical conditions that is contrary to our general approach, 2. If the FS is marginally fast cooling, the RS is not very deep in the fast cooling regime, and should show a cooling break near the FS peak photon energy, below which $1+b_1=4/3$ is expected in the total spectrum. 3. We use an approximated way of calculating the radiation in this regime: we expect the thin shell assumption to break and thus effects of a finite size emission region to appear, which are bound to modify emission signatures. Such effects will be the focus of a following article dedicated to the exploration of the cooling regimes.

\section{Conclusions}\label{sec:ccl}
We have presented a fully spherical, self-consistent numerical approach of the internal shocks model for the prompt GRB emission, showing the impact of the spherical hydrodynamics effects on emission, even for crossing radii lower than twice the initial collision radius. Our basic two-shell collision model describes a single spike in the prompt GRB light curve, which usually consists of multiple pulses that in this picture correspond to multiple collisions. After calibrating our methods against the analytical results obtained in \cite{rahaman2024internal, rahaman2024prompt} using a planar approach for hydrodynamics and approximate spherical effects for the emission, we numerically extend their conclusions to the fully spherical case while keeping the assumption of an infinitely thin emitting shell. We present the structural and dynamical differences of shocked shells in spherical geometry and highlight the non-negligible effects of the evolution in Lorentz factors and shock strengths on the emission over the shock crossing time. In particular, the quantities associated with each shock front evolve at a different rate due to the differences in shock strengths. We then expect that following the spectral properties of a burst over time will allow to distinguish this two-zones emission model from a one-zone model. While we assumed equal values of the microphysical parameters $(\epsilon_\mathrm{e}, \epsilon_\mathrm{B}, \xi_\mathrm{e}, p)$ between the two shock fronts, any diversity in those parameters will most likely add diversity to the obtained emission between the two shock fronts. 

We produce light curves and (time-integrated) spectra with two different spectral shapes, the smooth Band function and the synchrotron broken power-law. While the Band function is a phenomenological model of the emission spectrum, the synchrotron broken power-law is physically motivated. Calculating flux with the latter results in less flux at lower frequencies, but conserves the general time evolution of the light curve. In both choices of spectral shape, our framework naturally obtains a doubly-broken power-law shape for the spectrum, shape that has been successfully used to fit prompt GRB data \citep[see e.g.][]{burgess2014time, oganesyan2017detection, ravasio2019evidence}. The validity of the results obtained with the rough spectral shape compared to using the empiric smooth Band function motivates the use of this simpler emission function to explore more complicated flows as well as other cooling regimes with different spectral shapes in further works while retaining physical accuracy. The value of the spectral slope between the two breaks varies significantly with the choice of initial conditions. While not totally consistent with the range of values obtained in \cite{ravasio2019evidence}, this is not a contradicting result. Those slopes are more dependent on the choices made for the fitting functions than the low- and high-energy part, and a better test of our model would be to perform fits of observed data from calculated flux across an expected parameter space.

While the effects of spherical geometry on the shocks hydrodynamics are not striking when comparing light curves and time-integrated spectra, they cannot be ignored when studying the evolution of spectral shape within a single spike in the prompt GRB light curve. In particular we obtain shapes that are more similar to the observed ones for a choice of GRBs presented in \cite{yan2024one}, and our inferred crossing radii and source activity time over off time to obtain such peak flux and frequency evolution are smaller by a factor of 2 and 3, respectively, when compared to the estimations done by \cite{rahaman2024prompt}.

This work relies on many highly idealized assumptions, especially the choice of homogeneous shells in both density and Lorentz factor. This assumption may be lifted to obtain more realistic emission signatures of internal shocks by simulating more physically-motivated flows. A striking example is the sharp edge in the peak flux vs peak frequency plot that originates from the sharp edge of the shells, emission stopping at once when the shock crosses. Such a sharp edge is not seen in observed GRBs \citep{yan2024one}). The exploration of internal shocks in higher dimensions is also expected to introduce additional effects. Studies on the external shock of GRB outflows (i.e. the one that form due to the interaction with the external medium) in 2D have shown the growth of Rayleigh-Taylor instabilities (RTI) significantly modifies the dynamics of the external reverse shock and causes its emission to peak at a later time \citep[e.g.][]{duffell2013rayleigh, ayache2022gamma}. The shear flow and/or turbulence near RTI fingers at the contact discontinuity may also be source of particle acceleration.

Additionally, the synchrotron spectra used in this work is obtained in the limiting case where all the non-thermal electron energy is radiated away in less than a numerical time step. This limits the emission region to a single numerical cell representing an infinitely thin shell, behind each shock front. This case assumes being very deep in the fast cooling regime behind both shock fronts, producing spectral slopes that do not agree to the low energy part of observed spectra fitted with a doubly-broken power-law, such as in \cite{ravasio2019evidence}. Their method obtains low-energy slopes indicative of slow cooling regime, which is naturally not reached in our assumption. A simple calculation by changing the FS emission to an approximate marginally fast cooling regime was not enough to obtain slopes in this range. But exploring the impact of other cooling regimes and extended emission zones in a consistent way using the capacity of \texttt{GAMMA} to evolve a non-thermal electron distribution could be a step towards solving this issue and will be the focus of a future work. Changing the microphysical parameters is one of the ways to explore other cooling regimes and try to obtain the observed photon indices. In \cite{bovsnjak2014spectral}, they find the best agreement to observations using varying microphysical parameters with shock conditions. Numerically, this could be achieved by locally computing the value of these parameters following recipes. \texttt{GAMMA} already proposes one such recipe to compute $p$, exploring its effects and implementing local evolution of other parameters may be pursued in a future work.

The implementation for the local electron distribution in \texttt{GAMMA} only features synchrotron and adiabatic cooling, but inverse Compton scatterings with the photons produced at both fronts are expected to modify the distribution in the context of GRB emission (see e.g. \citealt{nakar2009klein} for a comprehensive reference, \citealt{Jacovich2021,mccarthy2024klein, pellouin2024very} for recent numerical implementations in the context of GRB afterglows). Implementing proper inverse Compton and synchrotron self-Compton is a challenging task that may be tackled in future works.

\begin{acknowledgements}
We thank Frédéric Daigne for the insightful discussions and the CRAL for its welcome. PB is supported by a grant (no. 2020747) from the United States-Israel Binational Science Foundation (BSF), Jerusalem, Israel, by a grant (no. 1649/23) from the Israel Science Foundation and by a grant (no. 80NSSC 24K0770) from the NASA astrophysics theory program. 
\end{acknowledgements}


\bibliographystyle{aa} 
\bibliography{biblio} 


\begin{appendix}

\section{Hydrodynamics in planar geometry}\label{app:planhydro}
In planar geometry, the quantities are constant across the two shocked regions. In particular the shocked material's proper speed, $u_3=u_2\equiv u_0$, can be conveniently determined in the rest frame of shell 1 (see Appendix B of R24a):
\begin{equation}
    u_{21} = u_{41}\left(\dfrac{2f^{3/2}\Gamma_{41}-f(1+f)}{2f(u_{41}^2+\Gamma_{41}^2)-(1+f^2)}\right)^{1/2}\;,\label{eqn:u21}
\end{equation}
where $u_{41}$ and $\Gamma_{41}$ are the proper velocity and Lorentz factor of shell 4 in the rest frame of shell 1. Eq.~(\ref{eqn:u21}) is transformed to the source frame by velocity transformation $u_0=\Gamma_{21}\Gamma_1(\beta_{21}+\beta_1)$. The density and pressure in the shocked shells are determined from the relative Lorentz factor between up- and downstream, $\Gamma_{34}$ for the RS and $\Gamma_{21}$ for the FS, using shock jump conditions. $\Gamma_{21}$ is derived from Eq.~(\ref{eqn:u21}), and $\Gamma_{34}$ is easily obtained by either velocity transformation or using the result $(\Gamma_{21}^2-1)=f(\Gamma_{34}^2-1)$ that applies to the collision of cold shells (see R24a). The shock front velocities $\beta_\mathrm{RS}$ and $\beta_\mathrm{FS}$ can be derived from these relative Lorentz factors (see Appendix C of R24a for the full derivation), and in turn one obtains the crossing times of both shocks:
\begin{align}
    &\beta_\mathrm{RS} = \frac{\beta_4-4\Gamma_{34}\left(\frac{u_0}{\Gamma_4}\right)}{1-4\Gamma_{34}\left(\frac{\Gamma_0}{\Gamma_4}\right)}\;, &\beta_\mathrm{FS} = \frac{\frac{1}{4\Gamma_{21}}\left(\frac{u_1}{\Gamma_0}\right)-\beta_2}{\frac{1}{4\Gamma_{21}}\left(\frac{\Gamma_1}{\Gamma_0}\right)-1}\;,\\
    &t_\mathrm{RS}=\dfrac{\beta_4t_{\mathrm{on},4}}{\beta_4-\beta_\mathrm{RS}}\;, &t_\mathrm{FS}=\dfrac{\beta_1t_{\mathrm{on},1}}{\beta_1-\beta_\mathrm{FS}}\;.\label{eqn:t_sh}
\end{align}

\begin{figure*}
    \centering
    \includegraphics[width=.32\textwidth]{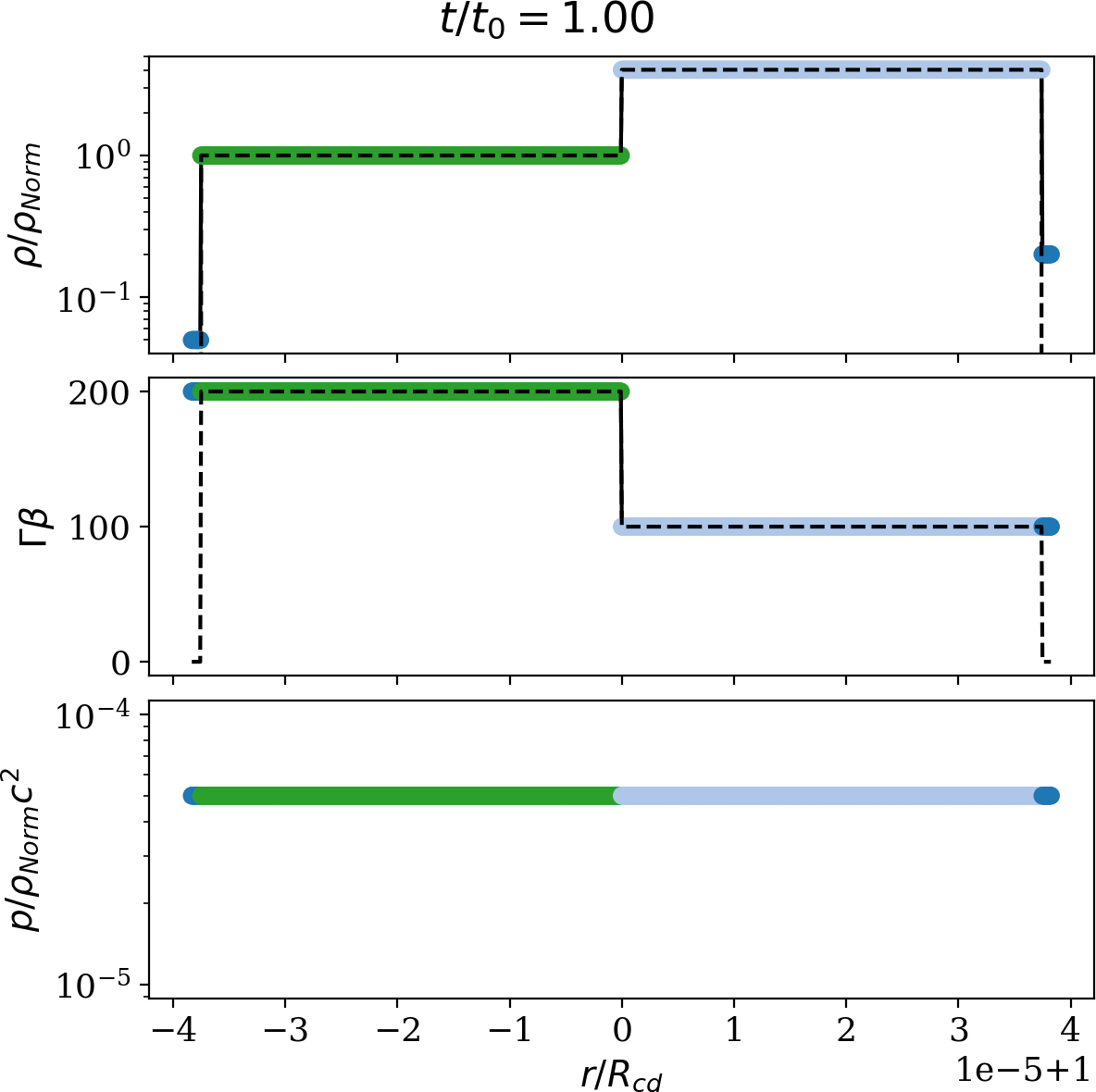} \includegraphics[width=.32\textwidth]{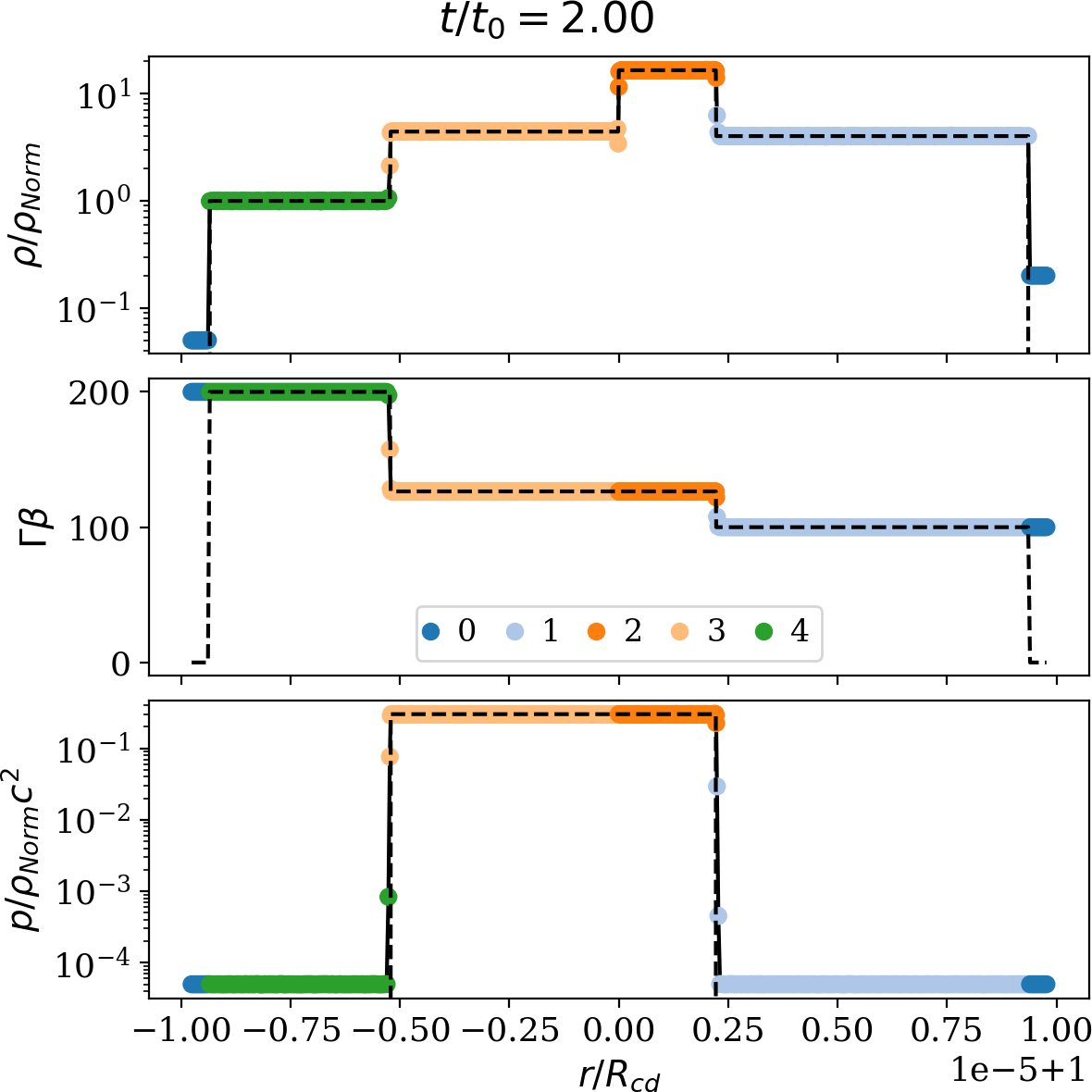} \includegraphics[width=.32\textwidth]{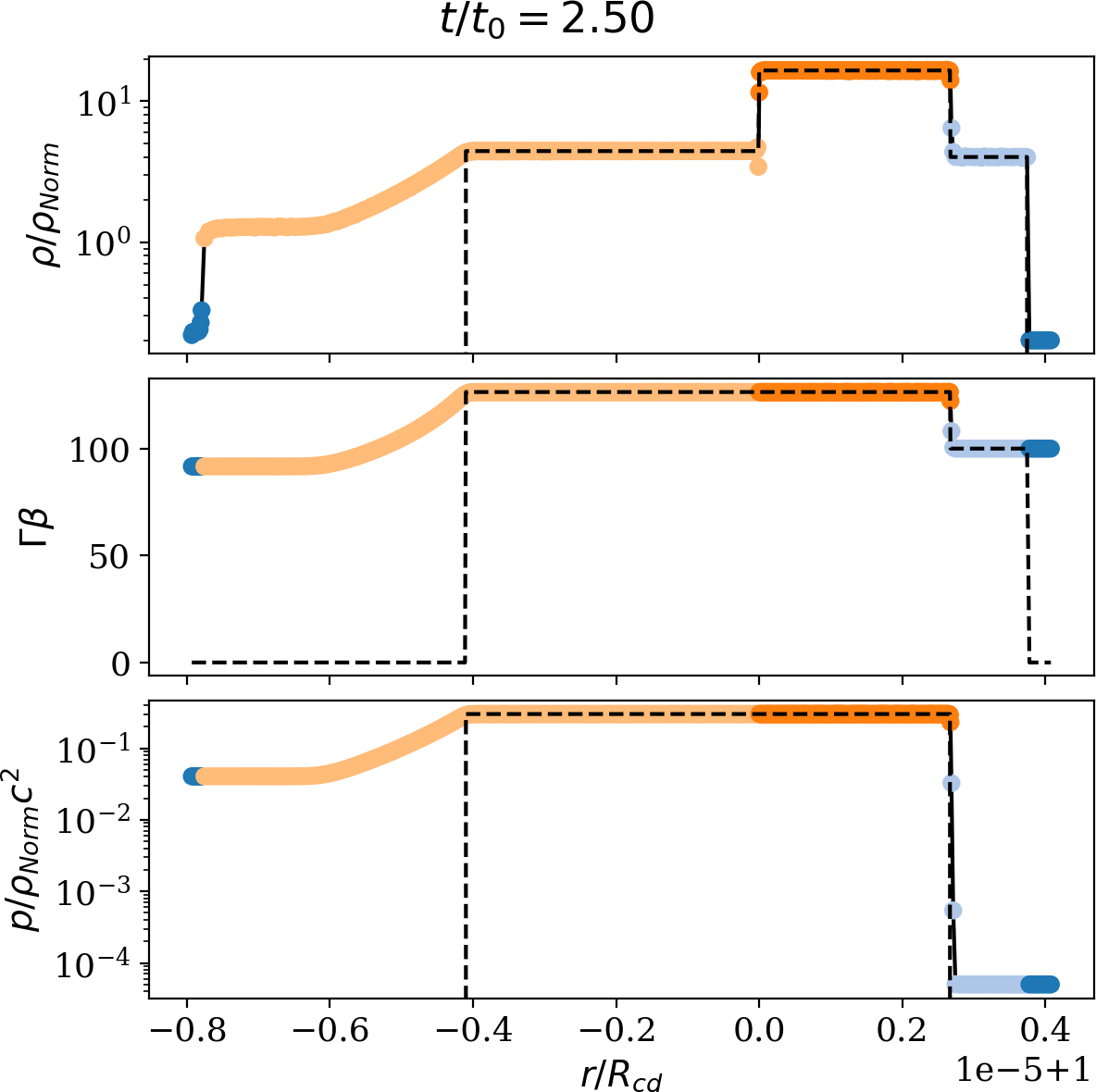}
    \caption{Snapshots of the fiducial planar run ($a_\mathrm{u}=2$, $E_{\mathrm{k}0,4}=E_{\mathrm{k}0,1}$, $t_{\mathrm{on},1}:t_{\mathrm{on},4}:t_\mathrm{off} = 1:1:1$.) at $t=t_0$ (left), before first shock crossing (center), and after shock crossing (right). Shells are color-coded according to the identified zones, the dashed lines show theoretical values. $t_0=2.7\times10^3$ s and $R_0 = 8\times10^{13}$ cm for this set of parameters.}
    \label{fig:cart_snaps}
\end{figure*}

To make sure our numerical methods fit the analytical framework of \cite{rahaman2024internal}, we ran the simulation described in \S\;\ref{sec1:setup} up to the point where each of the shocks have crossed their respective shells. Fig.~\ref{fig:cart_snaps} shows three snapshots of our fiducial simulation: (a) the initial conditions, (b) $t_0$ after the collision, (c) 1.5 $t_0$ after the collision. The dashed black lines show the expected values using the analytical results from R24a for the same set of parameters. The zones are determined with a combination of the passive tracers, the position of the shocks identified by \texttt{GAMMA}, and an interface detection algorithm based on peak detections in the gradient of the hydrodynamical variables to identify the rarefaction wave front when no shock is present in a given shell. We see a very good agreement for both the hydrodynamical variables and the position of the interfaces of both shocks and the rarefaction wave after the first shock crossing. This interface analysis is completed with a method to recover the shock front velocities by detecting when the front enters a cell for the first time, and assuming the front is at the interface between this cell and its already-shocked neighbor at the half time-step.

The values across the shocked region remain constant in time with very good accuracy compared to the analytical results: the proper speed $u$ matches its expected value with a precision of the order $10^{-5}$, and derived quantities such as the shock strength $\Gamma_\mathrm{ud}-1$ matches to within a few times $10^{-4}$. We also recover the analytical values for the shock crossing times $t_\mathrm{RS}$ and $t_\mathrm{FS}$ with an accuracy of $10^{-3}$, consistent with a systematic error of a few time steps due to the finite width of the numerical shock.

\section{Flux calculation in the power-law approximation}
\subsection{Analytical flux}\label{app:flux_gen}
We present here the formalism for the analytical calculation of the flux received at observer time $T$ and frequency $\nu$ from a range of emitting radii. From Eq.~(\ref{eqn:Fnu_GG09}) we see a single pulse emitted at a single time and radius in the source frame will be seen over a range of observed times. This means that several source times and radii contribute to a single observer time, defining a surface called equal arrival time surface (EATS) as the locus of all emission points from which photons will arrive at the same time to the observer. In R24b, itself expanding on previous works \citep[e.g.][]{Granot-05,Granot+08,genet2009realistic}, the authors derive the observed flux coming from a shock front propagating with Lorentz factor $\Gamma_\mathrm{sh}=\Gamma_\mathrm{sh,0}(R/R_0)^{-m/2}$, emitting between radii $R_0$ and $R_f=(1+\Delta R/R_0)R_0$ at a peak frequency $\nu'_\mathrm{p}=\nu'_0(R/R_0)^d$ and with peak luminosity $L'_{\nu'_\mathrm{p}} = L'_0(R/R_0)^a$. As quantities introduced in this subsection are relative to a single shock front, we will drop here the subscript $RS$ (respectively $FS$) for the sake of readability. Additionally in R24, compared to GG09, it is the shocked material traveling with Lorentz factor $\Gamma = g\Gamma_\mathrm{sh}$ that is considered to be the source of emission, meaning that while the shock front itself serves to determine the location of the emitting surface, it is the emitting material that will be considered for the Doppler factor.

The flux is calculated against normalized time $\tilde{T}$ defined as:
\begin{equation}
    \tilde{T}=\frac{T-T_{\mathrm{ej},0}}{T_0}\;,\label{eqn:barTRS}
\end{equation}
where $T_{\mathrm{ej},0}$ and $T_0$ are the effective (observed) ejection time and angular time as defined Eq.~(\ref{eqn:obsT}) for the shock front at $R_0$ the initial collision radius:
\begin{align}
    T_0 &= (1+z)\left(\frac{1-\beta_\mathrm{sh,0}}{\beta_\mathrm{sh,0}}\right)\frac{R_0}{c} \approx (1+z)\frac{R_0}{2c\Gamma_\mathrm{sh,0}^2}\;,\\
    T_{\mathrm{ej},0} &= (1+z)\left(t_0-\frac{R_0}{\beta_\mathrm{sh,0}c}\right)\;.
\end{align}
We do not provide here the details of the EATS and intermediate variables definitions, and instead refer again to appendix F of R24b for full details of the calculation. We introduce the normalized radius $y\,(\tilde{T})\equiv R/R_L(\tilde{T})$ such that the flux is obtained by integrating from $y_\mathrm{min}$ to $y_\mathrm{max}$ defined as:
\begin{align}
    y_\mathrm{min}(\tilde{T}) &= \begin{cases}
        1\quad&\text{if }\tilde{T}\leq 1\;,\\
        \tilde{T}^{-1/(m+1)}\quad&\text{if }\tilde{T}\geq 1\;,
    \end{cases}\label{eqn:ymin}\\
    y_\mathrm{max}(\tilde{T}) &= \begin{cases}
        1\quad&\text{if }\tilde{T}_f\leq 1\;,\\
        \tilde{T}_f^{-1/(m+1)}\quad&\text{if }\tilde{T}_f\geq 1\;,
    \end{cases}\label{eqn:ymax}
\end{align}
where $\tilde{T}_f=(1+\Delta R/R_0)^{m+1}$. The Doppler factor is also expressed as a function of $y$:
\begin{equation}
\begin{split}
    \delta = (1+z)\frac{\nu}{\nu'}=
    \frac{1+z}{\Gamma(1-\beta\mu)}
\approx(1+z)\frac{2(m+1)g\Gamma_\mathrm{sh,0}y^{\frac{m}{2}}\tilde{T}^{\frac{-m}{2(m+1)}}}{g^2y^{-1}+(m+1-g^2)y^m}\;.\label{eqn:EATS_D}
\end{split}
\end{equation}
The received flux at normalized observer time $\tilde{T}$ for any given comoving luminosity $L'_{\nu'}(y)$ can be expressed as:
\begin{equation}
    F_\nu (\tilde{T})= \frac{1+z}{8\pi d_L^2} \int_{y_\mathrm{min}}^{y_\mathrm{max}} \left|\frac{{\rm d}\mu}{{\rm d}y}\right|\delta^3L'_{\nu'}(y){\rm d}y.\label{eqn:Fnu_th_basic}
\end{equation}
With the change of variable:
\begin{equation}
    \frac{{\rm d}\mu}{{\rm d}y} = \frac{y^{-2}(1+my^{m+1})}{2(m+1)\Gamma_{\rm sh,0}^2}\tilde{T}^{m/(m+1)},
\end{equation}
and using Eq.~(\ref{eqn:EATS_D}) in \ref{eqn:Fnu_th_basic}:
\begin{equation}
\begin{split}
    F_\nu(\tilde{T}) = \frac{(1+z)g^3\Gamma_\mathrm{sh,0}\tilde{T}^{\frac{-m}{2(m+1)}}}{2\pi d_L^2}\int_{y_\mathrm{min}}^{y_\mathrm{max}}\mathrm{d}y\,L'_{\nu'}(y)y^{-1-\frac{m}{2}}\quad\quad\quad\ \ \\ \quad\quad\quad\ \ \times\left[1+g^2\frac{y^{-(m+1)}-1}{m+1}\right]^{-2}\left(\frac{1+my^{m+1}}{g^2+(m+1-g^2)y^{m+1}}\right)\;.\label{eqn:Fnu_th_general}
\end{split}
\end{equation}
In the general case, we will chose power-law scalings for the luminosity and peak frequency with radius:
\begin{align}
    L'_{\nu'} &= L'_0\left(\frac{R}{R_0}\right)^a S\left(\frac{\nu'}{\nu'_p}\right) = L'_0 \tilde{T}^a y^a S\left(\frac{\nu'}{\nu'_p}\right)\;,\\
    \frac{\nu'}{\nu'_p} &= \tilde{T}^{-d}y^{m/2-d}\left[1+g^2\frac{(y^{-(m+1)}-1)}{m+1}\right]  \frac{\nu}{\nu_0}\;.
\end{align}

\subsection{Numerical flux normalization}\label{app:flux_norm}
To obtain the appropriate flux from a numerical cell corresponding to the infinitely thin shell approximation used in R24b, we compare the isotropic energy obtained by integrating Eq.~(\ref{eqn:Fnu_GG09}) to the equivalent integration of the flux formula from \cite{de2012gamma} for the contribution from a single computational cell, which for an isotropic comoving emissivity per unit volume and frequency, $P'_{\nu',\mathrm{jk}}$, reads:
\begin{equation}
\begin{split}
&\Delta F_{\nu,\mathrm{ijk}} = \frac{(1+z)^2}{4\pi d_\mathrm{L}^2} \frac{\Delta V_\mathrm{ijk}^{(4)}}{\Delta T_\mathrm{obs,i}} \frac{P'_{\nu',\mathrm{jk}}}{\Gamma_\mathrm{jk}^2 (1-\beta_\mathrm{jk}\cos\theta_\mathrm{jk})^2}\;,\\
    &\quad\quad\quad\text{for }\abs{T_\mathrm{obs,z,i} - t_\mathrm{jk} + \frac{r_\mathrm{jk}\cos\theta_\mathrm{jk}}{c}} < \frac{\Delta T_\mathrm{obs,z,i}}{2}\;.
\end{split}\label{eqn:dFnu_dC12}
\end{equation}
Here $T_\mathrm{obs,z}=T_{\rm obs}/(1+z)$ and $\nu_z=(1+z)\nu$ are measured in the source's cosmological frame, as is the (Lorentz invariant) 4-volume $\Delta V^{(4)}$. In this section, subscript i refer to observer time bins over which flux is calculated, subscript j to the simulation time step, and subscript k to the numerical cell within data file j. This approach attributes all of the contribution from any given numerical 4D cell jk to a single observer time interval i. This approximation relies on sufficiently fine spatial (and temporal) resolution and on the Doppler factor not significantly varying within the cell, due to the variation of its velocity direction with respect to the direction to the observer. However, our simulation is spatially 1D: the Doppler factor significantly varies within each numerical cell of lab-frame volume $\Delta V^{(4)}_{\rm jk} = 4\pi r^2\Delta r$ and contributes to different observed time bins. We thus divide 1D cells further into subcells along the $\theta$ coordinate in which the Doppler factor can be considered constant.

The observed flux depends only on $\mu=\cos\theta$ through the Doppler factor and the photon arrival time but not on the $\phi$ coordinate as we calculate flux along the outflow axis. Thus each subdivision of volume $\Delta V^{(4)}_\mathrm{ijk}=2\pi r^2_\mathrm{jk}\Delta r_\mathrm{jk}\Delta\mu_\mathrm{i}\Delta t_\mathrm{j}$ contributes to the time bin centered on $T_{\rm obs,i}= (1+z)(t_{\rm j} - r_{\rm jk}\mu_{\rm i}/c)$ of width $\Delta T_\mathrm{obs,i}=(1+z)(r_{\rm jk}/c)\Delta\mu_i$. Recalling $E_{\rm iso}=\int d\nu_z\int dT_\mathrm{obs,z} L_{\rm\nu_z,iso}=\left(4\pi d_\mathrm{L}^2/(1+z)\right)\int dT_{\rm obs}\int d\nu F_\nu(T_{\rm obs})$, the isotropic energy for a spherical cell reads:
\begin{equation}
\begin{split}
E_\mathrm{iso,jk}^\mathrm{(num)} &= \frac{4\pi d_\mathrm{L}^2}{1+z}\int d\nu\,\sum_{\mathrm{i}}\Delta T_\mathrm{obs, i}\Delta F_{\nu,\mathrm{ijk}}\\
&= \int d\nu_z\,\sum_{\mathrm{i}}\Delta V_\mathrm{ijk}^{(4)}\frac{P'_{\nu',\mathrm{jk}}}{\Gamma_\mathrm{jk}^2 (1-\beta_\mathrm{jk}\mu_\mathrm{i})^2}\\
    &= 2\pi r^2_\mathrm{jk}\Delta r_\mathrm{jk} \Delta t_\mathrm{j} \,\frac{P'_{\nu'_\mathrm{m,jk}}}{\Gamma_\mathrm{jk}^{2}} \int_0^\infty d\nu_z \int_{-1}^{1} d\mu (1-\beta\mu)^{-2} S\left(\frac{\nu'}{\nu'_\mathrm{m,jk}}\right) \\
    &= \Delta V_\mathrm{jk}^{(4)}(1-\beta_\mathrm{jk})\Gamma_\mathrm{jk}^2 \nu_\mathrm{m,z,jk} P'_{\nu'_\mathrm{m,jk}}\mathcal{A}
    \cong \frac{1}{2}\Delta V_\mathrm{jk}^{(4)}\nu_\mathrm{m,z,jk}P'_{\nu'_\mathrm{m,jk}}\mathcal{A}\;,
\end{split}
\end{equation}
having changed variables to $x\equiv\nu'/\nu'_\mathrm{m} =\nu(1-\beta\mu)/\left[\nu_\mathrm{m}(1-\beta)\right] = \nu_z(1-\beta\mu)/\left[\nu_\mathrm{m,z}(1-\beta)\right]$ and denoting $\mathcal{A}=\int S(x)\mathrm{d}x$. In \cite{de2012gamma} the flux at a fixed observed time bin was calculated by summing over j and k. To obtain the isotropic energy we performed the opposite, i.e. fix j and k and sum over i (and integrated over frequency). Performing the equivalent integration on a single shell in the infinitely thin shell approximation \citep{genet2009realistic} reads:
\begin{equation}
\begin{split}
E_{iso}^{(th)} &= \frac{4\pi d_\mathrm{L}^2}{1+z}\int dT_\mathrm{obs}\int d\nu\, F_\nu 
    &=\tilde{L}_{\nu_\mathrm{m}}T_\theta
    \int \frac{d\tilde{T}}{\tilde{T}^2}\int d\nu 
     S\left(\frac{\nu}{\nu_\mathrm{m}} \tilde{T}\right)=\frac{1}{2}\nu_\mathrm{m,z} \tilde{L}_{\nu_\mathrm{m,z}} T_{\theta,z} \mathcal{A}
    \;,
\end{split}
\end{equation}
using change of variable $x \equiv (\nu/\nu_\mathrm{m})\tilde{T}$. Finally, we obtain the equivalent numerical luminosity for simulation cell k of data file j in the infinitely thin shell approximation:
\begin{equation}
    \tilde{L}_{\nu_\mathrm{m,z},\mathrm{jk}} = \frac{\Delta V^{(4)}_\mathrm{jk}P'_{\nu'_\mathrm{m},\mathrm{jk}}}{T_{\theta,z,\mathrm{jk}}} = \frac{\Delta E'_{\nu'_\mathrm{m},\mathrm{jk}}}{T_{\theta,z,\mathrm{jk}}}\;.
\end{equation}
Here $\Delta E'_{\nu'_\mathrm{m},\mathrm{jk}}$ is the energy emitted per unit frequency in the comoving frame by the 4-dimensional cell jk. Going further with the infinitely thin shell approximation, we can assume all the energy given to the accelerated electrons is radiated over a time smaller than the numerical timestep, and we write:
\begin{equation}\label{eqn:Lnorm}
    \tilde{L}_{\nu_\mathrm{m,z},\mathrm{jk}} = \frac{\Gamma_\mathrm{jk}\Delta V^{(3)}_\mathrm{jk}\epsilon_\mathrm{e}e'_\mathrm{jk}}{W(p)\nu'_\mathrm{m,jk}T_{\theta,z,\mathrm{jk}}}\;.
\end{equation}

\subsection{Flux calibration}\label{app:flux_calib}
While the general analytical formula for any scaling indices $a, d, m$ was not obtained and the integral in Eq.~(\ref{eqn:Fnu_th_general}) must be calculated numerically, two special cases with a simpler integral form are of interest to us. We use the first case $a, d, m = 0, 0, 0$ to calibrate the post-processing methods in a consistent way from our fully planar simulation before extending them to the physically-relevant spherical geometry. With these simplifying assumptions, the observed flux is:
\begin{equation}
    F_\nu(\tilde{T}\geq1) = \frac{(1+z)g^3\Gamma_\mathrm{sh}L'_0}{2\pi d_\mathrm{L}^2} \int_{y_\mathrm{min}}^{y_\mathrm{max}}\frac{\mathrm{d}y\,y^{-2}S\left(x\right)}{\left[1+g^2(y^{-1}-1)\right]^3}\;.
\end{equation}
With
\begin{align}
    x &= \dfrac{\nu'}{\nu'_0} = \left[1+g^2(y^{-1}-1)\right]\tilde{\nu}\;,\\
    \tilde{\nu} &= \nu/\nu_0\;,\\
    \tilde{T} &= \frac{T-T_\mathrm{ej}(R_0, \beta_\mathrm{sh})}{T_{\theta}(R_0, \Gamma_\mathrm{sh})})\equiv1+\bar{T}\;,
\end{align}
and the integration limits $y_\mathrm{min}$, $y_\mathrm{max}$ are defined in Eqs.~(\ref{eqn:ymin})-(\ref{eqn:ymax}). Choosing x as the new integration variable, we obtain:
\begin{equation}
    F_\nu = \frac{(1+z)2\Gamma_0 L'_0}{4\pi d_\mathrm{L}^2}\left(\frac{\nu}{\nu_0}\right)^2\int_{x_\mathrm{l}}^{x_\mathrm{u}}\mathrm{d}x\,x^{-3}S(x)\;.\label{eqn:Fnu_th000}
\end{equation}
The upper and lower integration limits are respectively $x_\mathrm{u} = x(y_\mathrm{min})$ and $x_\mathrm{l} = x(y_\mathrm{max})$, and the integral is performed numerically using the \texttt{quad} method from \texttt{scipy.odeint}\footnote{An analytic solution of the integral exist for both spectral shapes, but in the case of the Band function this form features the incomplete Gamma function, which is not implemented in Python for negative values of the first parameter. See Appendix F in R24b.}. We present in Fig. \ref{fig:cart_nuFnu} light curves at $\tilde{\nu}=10^{-1}$ and instantaneous spectra at $\bar{T}=1$ obtained from our fiducial run in planar geometry, compared to the analytical flux obtained analytically with Eq.~(\ref{eqn:Fnu_th000}). The figure presents the observed flux using either the smooth Band function (top row) or a broken power-law (bottom row). Depending on the local spectral shape, the relative contributions of both shock to the light curve between their respective peak frequencies may vary, also causing a change in the peak flux at this intermediary frequency range, but the overall shape of the light curve obtained by the sum of both contributions do not. This difference between peak frequencies is better shown in the instantaneous spectrum where the shape of the curve between the peaks shows more variation with spectral shape. Our numerical fluxes agree to a good degree of precision with their analytical counterparts.

\begin{figure}
    \centering
    \resizebox{\hsize}{!}{\includegraphics[width=\linewidth]{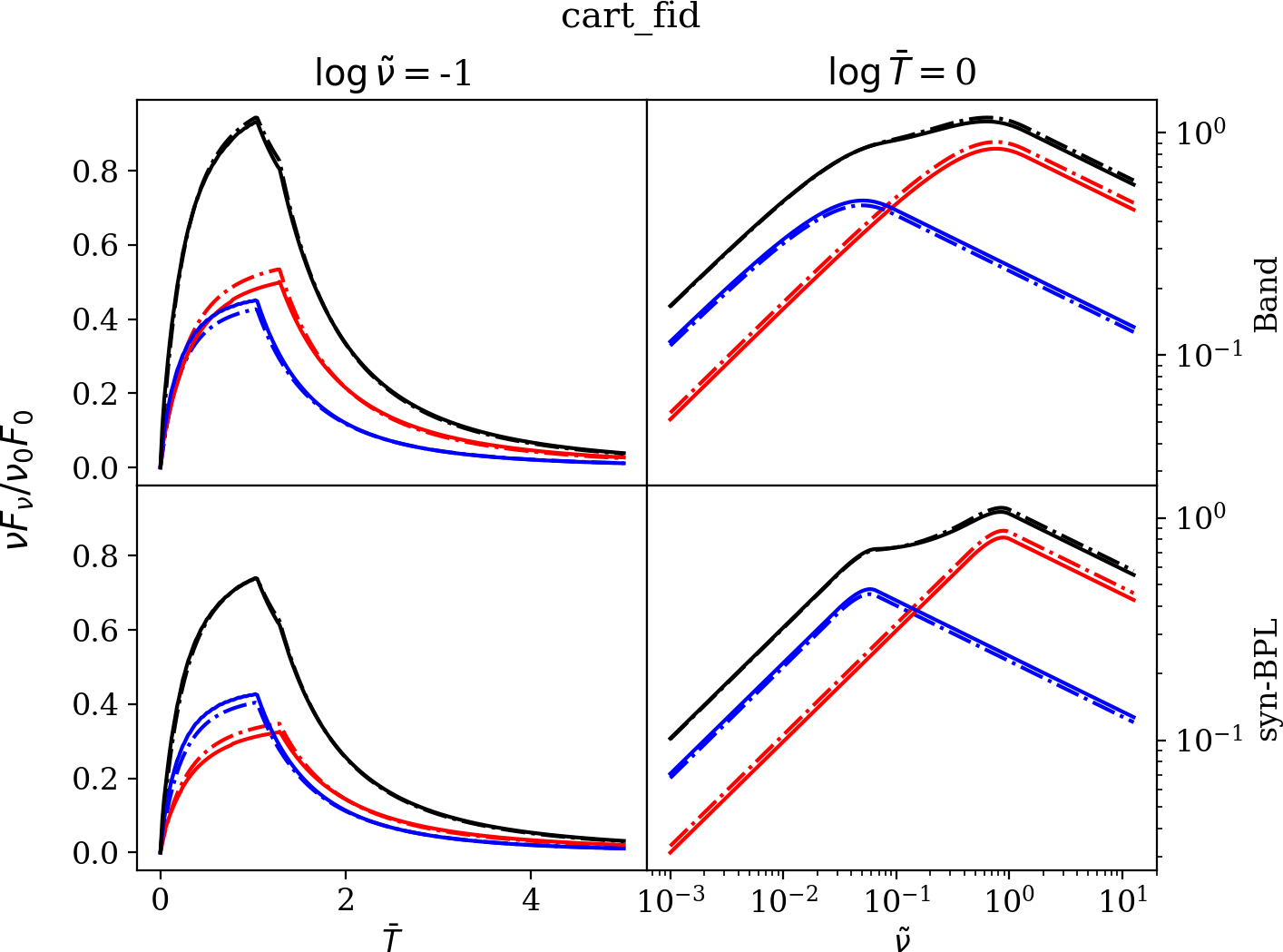}}
    \caption{Light curves (left column) and instantaneous spectra (right column) obtained from the fiducial run in planar geometry (full lines) compared to the analytical expectations (dash-dotted lines). Normalizations of $\bar{T}$ and $\tilde{\nu}$ are defined at the RS. The top row display the result for a Band spectral shape, and in the bottom the synchrotron broken power-law. The respective contributions of each shock to the total observed flux (in black) are given in red for the RS and blue for the FS.}
    \label{fig:cart_nuFnu}
\end{figure}

\end{appendix}
\end{document}